\documentclass[10pt,twocolumn,twoside]{IEEEtran}
\usepackage{amsmath,amssymb,amsfonts,mathrsfs,mathtools,bm, bbm, dsfont,mathrsfs,amsthm,blkarray}
\usepackage[dvipdfm]{graphicx}
\usepackage[dvipsnames]{xcolor}

\usepackage{epsfig,epsf,psfrag,latexsym}

\usepackage{booktabs}
\usepackage{multirow,multicol}

 \usepackage{orcidlink}

\PassOptionsToPackage{
  breaklinks,
  colorlinks,
  linkcolor=MidnightBlue,
  anchorcolor=MidnightBlue,
  citecolor=MidnightBlue,
  urlcolor=MidnightBlue
}{hyperref}

\usepackage{hyperref}
\usepackage{cite}

\newcommand*{\QED}{\hfill\ensuremath{\square}}

 \def\old#1{}    

\usepackage{bm}

\def\nn{\nonumber}
\def\beq{\begin{equation}}
\def\eeq{\end{equation}}
\def\bea{\begin{eqnarray}}
\def\eea{\end{eqnarray}}
\def\ba{\begin{array}}
\def\ea{\end{array}}

\def\bitem{\begin{itemize}}
\def\eitem{\end{itemize}}
\def\ben{\begin{enumerate}}
\def\een{\end{enumerate}}

\def\etal{{\it et al. \/}}

\def\ie{{\it i.e.,\ \/}}

\def\tcb{\textcolor{blue}}

\newcommand{\mbbE}{\mathbb{E}}

\newcommand{\mbbR}{\mathbb{R}}

\newcommand{\Fmsc}{\mathscr{F}}

\newcommand{\Smsc}{\mathscr{S}}

\def\mubf{\hbox{\boldmath$\mu$\unboldmath}}
\def\nubf{\hbox{\boldmath$\nu$\unboldmath}}

\def\pibf{\hbox{\boldmath$\pi$\unboldmath}}

\def\omegabf{\hbox{\boldmath$\omega$\unboldmath}}

\def\dbf{{\bm d}}
\def\ebf{{\bm e}}
\def\fbf{{\bm f}}
\def\gbf{{\bm g}}
\def\hbf{{\bm h}}

\def\pbf{{\bm p}}

\def\xbf{{\bm x}}
\def\ybf{{\bm y}}

\def\xbf{{\bm x}}
\def\ybf{{\bm y}}

\def\Cbf{{\bm C}}
\def\Dbf{{\bm D}}

\def\Sbf{{\bm S}}

\newcommand{\beqa}{\begin{eqnarray}}
\newcommand{\eeqa}{\end{eqnarray}}
\newcommand{\beqan}{\begin{eqnarray*}}
\newcommand{\eeqan}{\end{eqnarray*}}

\newcounter{l1}
\newcounter{l2}
\newcounter{l3}
\newcommand{\bdotlist}{\begin{list}{$\bullet$}{}}
\newcommand{\bboxlist}{\begin{list}{$\Box$}{}}
\newcommand{\bbboxlist}{\begin{list}{\raisebox{.005in}{{\tiny
$\blacksquare$ \ \ }}}{}}
\newcommand{\bdashlist}{\begin{list}{$-$}{} }
\newcommand{\blist}{\begin{list}{}{} }
\newcommand{\barablist}{\begin{list}{\arabic{l1}}{\usecounter{l1}}}
\newcommand{\balphlist}{\begin{list}{(\alph{l2})}{\usecounter{l2}}}
\newcommand{\bAlphlist}{\begin{list}{\Alph{l2}.}{\usecounter{l2}}}
\newcommand{\bdiamlist}{\begin{list}{$\diamond$}{}}
\newcommand{\bromalist}{\begin{list}{(\roman{l3})}{\usecounter{l3}}}

\newtheorem{theorem}{Theorem}
\newtheorem{lemma}{Lemma}
\newtheorem{proposition}{Proposition}

\begin{document}

\title{Wholesale Market Participation via Competitive DER Aggregation}

\author{Cong Chen$^{\ast}$\orcidlink{0000-0002-0547-2889}
       \quad
        Ahmed S. Alahmed$^{\ast}$\orcidlink{0000-0002-4715-4379}
        \quad
        Timothy D. Mount
        \quad
         Lang~Tong\orcidlink{0000-0003-3322-2681}
\thanks{\scriptsize Cong Chen ({\tt{\tcb{Cong.Chen@dartmouth.edu}}}) is with Thayer School of Engineering, Dartmouth College, USA. Ahmed S. Alahmed ({\tt{\tcb{alahmad@kfupm.edu.sa}}}) is with the Electrical Engineering Department at King Fahd University of Petroleum and Minerals, KSA. Timothy D. Mount ({\tt{\tcb{tdm2@cornell.edu}}}) is with the Dyson School of Applied Economics and Management, Cornell University,  USA. Lang Tong ({\tt{\tcb{lt35@cornell.edu}}}) is with the School of Electrical and Computer Engineering, Cornell University,  USA. ({$^\ast$\em Corresponding authors})}
\thanks{\scriptsize This work is supported in part by the National Science Foundation under Awards 2412776 and 2419622, the Power Systems Engineering Research Center (PSERC) Research Project M-46, and the Stanford Energy Postdoctoral Fellowship. Ahmed S. Alahmed acknowledges support from the Deanship of Research at KFUPM under project EC251013.}
\thanks{\scriptsize A preliminary version of this work was presented at the 56th Hawaii International Conference on System Sciences (HICSS) \cite{chen22competitive}.}
}

\maketitle

\begin{abstract}
We consider the aggregation of distributed energy resources (DERs), such as solar PV, energy storage, and flexible loads, by a profit-seeking aggregator participating directly in the wholesale market under distribution network access constraints. We propose a competitive DER aggregator (DERA) model that directly controls local DERs to maximize its profits, while ensuring each aggregated customer gains a surplus higher than their surplus under the regulated retail tariff. The DERA participates in the wholesale electricity market as virtual storage with optimized generation offers and consumption bids derived from the proposed competitive aggregation model.  Also derived are DERA’s bid curves for the distribution network access and DERA's profitability when competing with the regulated retail tariff. We show that, with the same distribution network access, the proposed DERA’s wholesale market participation achieves the same welfare-maximizing outcome as when its customers participate directly in the wholesale market. Extensive numerical studies compare the proposed DERA with existing methods in terms of customer surplus and DERA profit. We empirically evaluate how many DERAs can survive in the competition at long-run equilibrium, and assess the impacts of DER adoption levels and distribution network access on short-run operations.
\end{abstract}

\subsubsection*{Keywords}

distributed energy resources and aggregation, demand-side management, direct control, distribution network, net energy metering, competitive wholesale market.

\section{Introduction}\label{sec:Intro}

We address open problems in the direct participation of distributed energy resource aggregators (DERAs) in the wholesale electricity market operated by regional transmission organizations and independent system operators (RTOs/ISOs), as envisioned by FERC order 2222  \cite{FERC20}. We focus on the aggregation strategy of a profit-seeking DERA, whose industrial, commercial, and residential customers have competing service providers, such as their incumbent regulated utilities.  Our main objective is to design aggregation strategies that allow the DERA to remain {\em profitable} while offering {\em competitive} aggregation services that attract customers and grant the DERA operational control of their DERs.\footnote{Direct DERA control over DERs is inspired by programs such as the Tesla Virtual Power Plant, which establishes a control protocol once customers opt into the aggregation service \cite{TeslaVPP}.} By competitive aggregation, we mean that the benefits of the DERA customers must be no less than those offered by electricity provider benchmarks.  An example of such a benchmark is the incumbent utility or a community choice aggregator (CCA) adopting net energy metering (NEM) tariffs, offering strong incentives to prosumers with behind-the-meter (BTM) DERs \cite{birk2017tso, AlahmedTong2022NEM, SCE_FERC2222:21presentation}. A major barrier to DERA's entrance to direct wholesale market participation is having an aggregation strategy and a participation model to make DER aggregation a profitable venture, while competitively attracting customers \cite{borenstein2021designing}. 

The technical challenge of designing a {\em profitable and competitive} DER aggregation is twofold. First, the DERA plays a dual role in the aggregation process: an energy supplier to its customers in the retail market and a producer/demand in the wholesale market. Its aggregation must consider retail competition, distribution network access limits, and its overall revenue adequacy. To this end, a DERA  needs to derive profit-maximizing bids/offers from its competitive aggregation. 

Second, competitive aggregation requires the DERA to offer more attractive pricing to its customers than the regulated tariff and shield them from the volatility of wholesale market prices. Examples of unstable pricing are two-part tariffs from Griddy \cite{Griddy} and Amber \cite{Amber} defined by the wholesale spot price and a connection charge. Although Griddy’s aggregation offered competitive pricing compared to regulated utility tariffs, its customers experienced a 100-fold price surge during the extreme winter storm Uri in 2021.


\subsection{Related Work}

FERC Order 2222 removes regulatory barriers for DERAs to participate in wholesale capacity, energy, and ancillary service markets. In this paper, we focus on DERA participation in the energy market, where aggregators submit quantity \cite{GaoAlshehriBirge:22} or price-quantity bids \cite{Alshehri&etal:20TPS}, and the ISO clears the market and issues dispatch signals. Our analysis centers on market efficiency, DERA profitability, and how aggregators directly control DERs \cite{TeslaVPP, AutoGrid23VPP} to follow dispatch signals. While this study focuses on energy markets, the proposed framework is also extensible to capacity and ancillary service markets.

The growing literature on DER aggregation and wholesale market participation models broadly falls into two categories.  One is through a distribution network optimization operated by a distribution system operator (DSO) \cite{manshadi2015hierarchical, Huo24TCNSDERprivacy}, an aggregation/sharing platform \cite{morstyn2018multiclass, Ferro22TCNSp2p}, or an energy coalition \cite{chakraborty2018analysis, Alahmed25TCNS}. For the most part, these works do not consider a profit-maximizing DERA's active participation in the wholesale market at the transmission grid. In particular, in \cite{manshadi2015hierarchical, Huo24TCNSDERprivacy, morstyn2018multiclass}, the DSO or an aggregation platform participates in the wholesale markets with the aggregated power, treating the transmission network and wholesale market as a balancing resource.
 
 Our approach belongs to the second category of DER aggregations, where profit-seeking DERAs aggregate both generation and flexible demand resources, participating directly in the wholesale market with bid/offer curves. To ensure secure distribution network operation, DERA obeys the allocated distribution network access limits \cite{ChenBoseMountTong23DERA}  (a.k.a. operating envelopes \cite{Alahmed25TCNS} or feasibility sets \cite{Lee19TCNS, low14TCNSconvex}), rather than considering the computationally expensive network power flow constraints. With the direct wholesale market participation supported by FERC Order 2222, this type of DER aggregation has the potential to improve the overall system efficiency and reliability.

Although the notion of competitive DER aggregation has not been formally defined,  two prior works have developed competitive aggregation solutions in \cite{chakraborty2018analysis,GaoAlshehriBirge:22}. In \cite{chakraborty2018analysis}, Chakraborty \etal consider DER  aggregation by a CCA, where the authors provide an allocation rule that offers its customers competitive benefits with respect to the regulated utility.

Most relevant to our work is the DERA's wholesale market participation method developed by Gao {\em et al.} \cite{GaoAlshehriBirge:22} where the authors consider a profit-seeking DERA aggregating BTM distributed generations (DGs) and offering its aggregated generation resources to the wholesale market.  In particular, their approach achieves a social surplus equal to that achievable by customers' direct participation in the competitive wholesale market.  In other words, their approach achieves the highest economic efficiency for aggregating DGs. A significant difference between \cite{GaoAlshehriBirge:22} and this paper is that we formulate a general competitive aggregation that includes the regulated utility. In achieving DERA's profit maximization, our aggregation and market participation are also different from \cite{GaoAlshehriBirge:22}.

The approach proposed in \cite{GaoAlshehriBirge:22} follows the earlier work in \cite{Alshehri&etal:20TPS} where a Stackelberg game-theoretic model is used. Both approaches assume that the DERA elicits prosumer participation with an optimized (one-part or two-part) tariff, and the prosumer responds with its quantity to be aggregated by the DERA. The real-time wholesale market price is reflected by the variable price in \cite{Griddy,Amber,GaoAlshehriBirge:22}. Such a variable price conveys low but volatile wholesale prices directly to customers. To protect customers from price spikes in real-time wholesale prices, methods like price caps \cite{NW23priceCap} were proposed.

\subsection{Summary of Results, Contributions, and Limitations}

In this paper, we substantially extend the DERA aggregation model in \cite{chen22competitive} to one that controls aggregated customers across multiple locations in distribution networks and incorporates security constraints on network injection and withdrawal limits. We further investigate the competitive aggregation impact on market efficiency, price stability, and long-run equilibrium.  

First, we propose a DER aggregation approach based on a constrained convex optimization that maximizes DERA surplus while providing higher customer surpluses than those offered by a competing aggregation model. In particular, we are interested in aggregation schemes that are competitive with the regulated utility rates such as various versions of regulated NEM rates,\footnote{NEM, analyzed in \cite{AlahmedTong2022NEM}, is an inclusive parametric tariff design that captures key features of the existing and proposed NEM tariffs.} with which a customer can make cost-benefit comparisons in her decision to become a customer of the DERA. We show that such a competitive DER aggregation, despite the aggregation involving real-time wholesale locational marginal price (LMP), has an energy cost no greater than the regulated NEM tariff. This implies that the proposed DER aggregation mechanism ensures price stability regardless of the volatility of the wholesale market  LMP, a property missing in Griddy's pricing model \cite{Griddy}. Meanwhile,  we establish the profitability of DERA when competing with NEM.

Second, we propose a virtual storage model for DERA's wholesale market participation compatible with the practical continuous storage facility participation considered by ISOs \cite{ISONE19CSF, ISONE21}. The DERA bidding curve is derived from the closed-form solution of the proposed  DERA  model. While the aggregation optimization explicitly involves wholesale market LMP, the virtual storage bidding curves do not require forecasting of LMP. We show that the proposed DERA wholesale market participation results in market efficiency equal to what is achievable when DERA's customers participate directly in the wholesale market.

Finally, we derive the benefit function of DERA over distribution network injection and withdrawal access limits. DERAs compete in the distribution network access auction proposed by \cite{ChenBoseMountTong23DERA} to acquire network access, and we empirically evaluate the number of surviving DERAs in the long-run competitive equilibrium. We also present a set of numerical results, comparing the surplus distribution of the proposed competitive aggregation solution with those of various alternatives, including the regulated utility. Among the significant insights gained are the higher social surplus, customer surplus, and DERA surplus achievable in the proposed competitive DERA model, when compared to other alternatives. 

A few remarks are warranted regarding the scope and limitations of this paper. First, the losses in distribution systems are not considered. {Incorporating line losses would introduce non-convexity and spatial coupling among prosumers, making theoretical analysis likely intractable. Instead, we implicitly consider line losses by assuming they are internalized in the design of access limits (\ie operating envelopes) and regulated utility tariffs.} Second, the contingency cases where DSO rejects cleared bids and offers from DERA for reliability concerns  \cite{FERC20} are neglected. Under the access limit allocation framework proposed in \cite{ChenBoseMountTong23DERA}, reliability concerns of DER aggregation are already satisfied under normal operating conditions. Lastly, although the proposed competitive aggregation offers higher benefits to DERA customers, it does so with a non-uniform payment, which might raise equity concerns. {Note that non-uniform payments have been proposed, and in some cases implemented, even for regulated utilities. See for example income-graduated-fixed charges \cite{borenstein2021designing}, PV-capacity-based charges \cite{NEM3_JointUtility}, and energy community  in \cite{Chen&Zhao&Low&Wierman:23TSG}. Future work can study mechanisms that limit the non-uniformity of the payment to the fixed charge part rather than the price.}

{\small
\begin{table}[htbp]
 \vspace{-0.3cm}
\caption{\small Major symbols}\label{tab:symbols} \vspace{-0.2cm}
\begin{center}
\vspace{-1em}
\begin{tabular}{ll}
\hline
$\dbf$:& consumption bundle of aggregated customers.\\
$\overline{\dbf},\underline{\dbf}$: & consumption bundle's upper and lower limits.  \\
$\overline{\Cbf},\underline{\Cbf}$: & distribution network injection and withdrawal limits.  \\
$g, G$:& BTM single and aggregated DG.\\
${\cal K}$:&  competitiveness constant for prosumer surplus.\\
$N$:& total number of prosumers.\\
$M$:& total number of  points of aggregation (PoAs).\\
${\cal N}_m$:& set of aggregated customers under the $m$-th PoA.\\
$\omega$: & payment function of the aggregated customer.\\
$\pi^+,\pi^-, \pi^0$: & import rate, export rate, and fixed charges of NEM. \\
$\pi$:& wholesale LMP.\\
$Q$:& aggregated net injection quantity of DERA.\\
$S_{\mbox {\tiny DERA}}$:& total surpluses of DERA and its aggregated prosumers.\\
$S_{\mbox{\tiny NEM}}$:& prosumers surplus under tariff NEM.\\
$\Smsc(\cdot)$:&  aggregated supply function.\\
$U(\cdot)$: & prosumer utility of energy consumption function.\\
$V(\cdot)$:& prosumer marginal utility function.\\
\hline 
\end{tabular}
 \vspace{-0.7cm}
\end{center}
\end{table}
}

\subsection{Paper Organization and Notations}
In $\S$\ref{sec:Framework}, we summarize the DER aggregation model and its main interactions. The problem of competitive DER aggregation is formulated in $\S$\ref{sec:Retail} where we derive the optimal aggregation solution. $\S$\ref{sec:Whosale} and $\S$\ref{sec:DERAinDSO} consider DERA's wholesale market participation and its bidding strategies in the distribution network access auction, respectively. Numerical simulations are presented in $\S$\ref{sec:CaseStudies}. Detailed mathematical proofs are in the appendix. 

A list of major designated symbols is shown in Table \ref{tab:symbols}. The notations used here are standard.   We use boldface letters for column vectors as in $\xbf=(x_1,\ldots, x_n)$. In particular, $\bm{1}$  is a column vector of all ones.   The indicator function is denoted by $\mathbbm{1}\{x_n \le y_n \}$, which equals 1 if $x_n \le y_n$, and 0 otherwise. $\xbf  \preceq \ybf$ means $x_n \le y_n, \forall n$. $\mbbR_+$ represents the set of all nonnegative real numbers. $[x]$ represents the set of integers from 1 to $x$, \ie $[x] := \{1,\ldots,x\}$. 

\section{DER Aggregation Model}\label{sec:Framework}

\begin{figure}[htbp]
    \centering
    \vspace{-0.3cm} \includegraphics[scale=0.7]{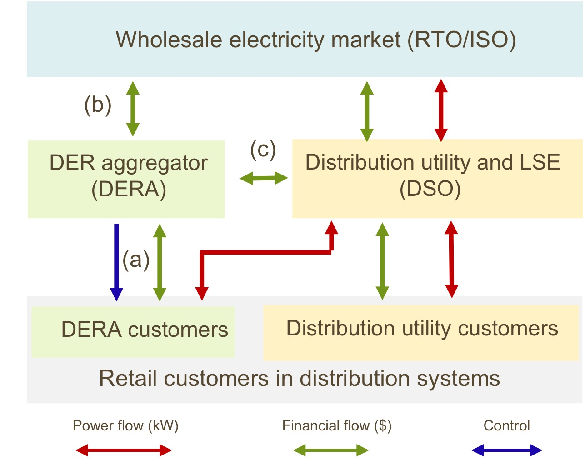}\vspace{-0.3cm}\caption{DERA model's physical and financial interactions. The red arrows show the bidirectional power flow, the green for the financial transactions, and the blue for the direct control signals. } \label{fig:DERAmodel}
    \vspace{-0.3cm}
\end{figure}

A DERA aggregates flexible resources from its customers and coordinates with the DSO for power delivery to the wholesale market operated by ISO/RTO. Following the DERA interaction model proposed in  \cite{ChenBoseMountTong23DERA}, we focus on the DERA-DSO-ISO/RTO interfaces (a)--(c), as shown in Fig. \ref{fig:DERAmodel}. Since a DERA uses DSO's physical networks for power delivery between its customers and the wholesale market, it is essential to delineate the financial and physical interactions at these interfaces. Below, we describe the three interfaces (a)--(c).
 
\textit{DERA and its customers at interface (a):} We assume the DERA aggregates BTM resources from retail market customers in the distribution network, where each customer alternatively has the option of being served by a regulated utility.  Under a single-bill payment model, each customer settles both consumption charges and production compensation with the DERA. The DERA deploys an energy management system that controls customers' BTM generation and flexible demand resources, such as rooftop PV, heat pumps, water heaters, and EV chargers. To remain competitive with the incumbent utility, the DERA must offer more attractive aggregation benefits; otherwise, customers will revert to the utility. These competitive aggregation benefits motivate customers to opt into the aggregation program, which includes an agreement granting the DERA direct control of their DERs. Customers retain the ability to operate their own DERs when desired, but the control agreement enables the DERA to coordinate DERs when participating in the wholesale market \cite{TeslaVPP}. Accordingly, the DERA optimizes and dispatches customers’ BTM resources and provides each customer with a cost–benefit comparison relative to the NEM benchmark offered by the utility.  See $\S$\ref{sec:Retail} for details.

\textit{DERA and RTO/ISO at interface (b):}   {We focus on DERA's participation in the energy market based on a {\em virtual storage} model compatible with the continuous storage facility model \cite{ISONE19CSF}. Specifically, the DERA submits offer/bid curves or self-scheduled quantities and may participate in both day-ahead and real-time markets; however, we focus on the real-time market with single-period optimization.\footnote{{See $\S$\ref{sec:Whosale} for bid/offer curves construction.}} In practice, DERAs may deploy their own DG and storage to mitigate aggregation uncertainties. Multi-period storage operation and multi-interval market models are beyond the scope of this paper.}

\textit{DERA and DSO at interface (c):} We consider the DERA-DSO coordination model in \cite{ChenBoseMountTong23DERA}, where DERA acquires access limits at distribution network buses operated by DSO. DERA's willingness to pay for network access is explained in $\S$\ref{sec:DERAinDSO}. Specifically, DERA secures injection and withdrawal access either through an access limit auction or a bilateral contract with the DSO. During real-time operation, DERA must aggregate DER from its customers in such a way that abides by the injection and withdrawal constraints set by the allocated access limits. That way, DERA's aggregation does not affect the operational reliability of DSO under nominal operating conditions, avoiding DSO intervention on ISO dispatch of DERA's aggregation.  

In summary, these three interactions at (a)--(c) establish our core framework for a DER aggregation model that is competitive, profit-making, and grid-aware.

\section{Competitive DER Aggregation}\label{sec:Retail}

This section formulates the optimal competitive aggregation and analyzes the properties of the optimal solution when competing with the incumbent utility's NEM. Our DER aggregation is built on the deregulated retail market. For example, in Texas and New York, customers can choose their electricity suppliers based on electricity rate and services. We consider prosumers owning all energy consumption and DG devices. After joining a DERA, prosumers grant device access to the DERA for measurements and direct controls.   
 \vspace{-0.3cm}
\subsection{Closed-Form Solution for Competitive DER Aggregation}\label{sec:closeform}

We consider a DERA aggregating customers over multiple points of aggregation (PoAs) in the distribution network.\footnote{For simplicity, we illustrate the single time interval aggregation model here and apply it to the multi-interval aggregation empirically in $\S$\ref{sec:EQ_MultiT}.} We define PoAs as the main buses with higher voltages in the distribution network, which can be identified from substation information. A diagram illustrating PoA is in Fig. 2 of \cite{ChenBoseMountTong23DERA}. With the DERA-DSO coordination method in \cite{ChenBoseMountTong23DERA}, DERA receives injection and withdrawal access limits at all PoAs, represented respectively by  
$$
\overline{\Cbf}:=(\overline{C}_{m}, m\in [M]),~~\underline{\Cbf}:=(\underline{C}_m, m \in[M]),
$$
where $\overline{\Cbf}, \underline{\Cbf} \in \mathbb{R}_+^M$, and $M$ denotes the total number of PoAs. Details about how the DERA coordinates with DSO to get distribution network injection and withdrawal limits are explained later in $\S$\ref{sec:DERAinDSO}. Thus, in the real-time operation, DERA must ensure that its aggregated power at the $m$-th PoA satisfies
\beq\label{eq:CLU}
-\underline{C}_m \le \sum_{n=1}^{{\cal N}_m}  (g_n-\mathbf{1}^\top \dbf_n) \le \overline{C}_m, \forall m \in [M],
\eeq
where $g_n\in \mathbb{R}_+$ represents the BTM DG output of the $n$-th aggregated customer and ${\cal N}_m$ the set of aggregated customers under the $m$-th PoA. Denote $N$ as the total number of DERA  customers, and mapping $\rho(n): [N] \rightarrow [M]$ such that $\rho(n)$ gives the index of PoA connecting  customer $n$,  then
\beq\label{eq:N}
{\cal N}_m := \{ n \in [N]~ |~ \rho(n) = m\}. 
\eeq 
$\dbf_n\in \mbbR_+^{K}$ is the consumption bundle of all customers, \ie
$$\dbf_n:=(d_{nk}, k\in [K]),$$
where $K$ denotes the total number of energy-consuming devices, including lamps, air-conditioners, washers/dryers, and heat pumps, for each customer $n \in [N]$. Customers set exogenous parameters $\underline{\dbf}_n,  \bar{\dbf}_n \in \mbbR_+^{K}$ as the minimum and maximum energy consumption limits of each device, \ie 
\beq\label{eq:DLU}
\underline{\dbf}_n \preceq  \dbf_n \preceq  \bar{\dbf}_n, \forall n \in [N].
\eeq
{\em Feasibility} of the DER aggregation requires that the distribution network access limits \eqref{eq:CLU} and consumption limits \eqref{eq:DLU} have a non-empty intersection at all times. Thus, we assume DERA acquires enough injection and withdrawal accesses such that
\beq\label{eq:FeasibleAssump}
\sum\limits_{n=1}^{{\cal N}_m}\sum\limits_{k=1}^K \underline{d}_{nk}-\underline{C}_m \le  \sum\limits_{n=1}^{{\cal N}_m} g_n \le \sum\limits_{n=1}^{{\cal N}_m}\sum\limits_{k=1}^K \overline{d}_{nk}+\overline{C}_m.
\eeq

To attain customers in the energy aggregation, DERA adopts the {\em ${\cal K}$-competitive constraint} in \eqref{eq:CompetiCons} to ensure that the surplus of each prosumer under aggregation is higher than the benchmark surplus ${\cal K}_n$ (e.g. surplus under the incumbent provider), \ie
\beq\label{eq:CompetiCons}
 U_n(\dbf_n)-\omega_n \geq {\cal K}_n, \forall n \in [N].
 \eeq
$ U_n(\dbf_n)$ is the $n$-th customer's utility of consuming $\dbf_n$. We assume the utility function is concave, nonnegative, nondecreasing, continuously differentiable, and additive  (\ie across the $K$ devices $U(\dbf)=\sum_{k=1}^K U_k(d_k)$). Here, the utility function is given; in practice, utility functions can be computed by parametric \cite{AlahmedTong2022NEM} or nonparametric \cite{varian82nonparametric} methods.

{\em ${\cal K}$-competitive constraint} is the criterion for a rational customer, seeking surplus maximization, to join a DERA. Otherwise, a rational prosumer has the incentive to leave DERA and switch to the benchmark service provider for a higher customer surplus.   More details about the benchmark prosumer surplus ${\cal K}_n$ are explained in $\S$\ref{sec:CNEM}.

To summarize, in real-time, the DERA solves for the  consumption bundle of all aggregated customers $\Dbf \in \mbbR_+^{N\times K}$ and their single-bill payments $\omegabf\in \mbbR^N$, defined by
$$\Dbf:=(\bm{d}_n, n \in [N]), ~~\omegabf:=(\omega_n, n \in[N])
$$
from the following convex profit maximization  
 \beq\label{eq:DERAsurplus_LnGPCC}
 \begin{array}{lcl}
\Pi(\overline{\Cbf}, \underline{\Cbf}):=&\underset{\omegabf,\Dbf}{\rm maximize}& \sum_{n=1}^N (\omega_n-\pi (\mathbf{1}^\top \dbf_n-g_n))\\
&  \mbox{subject to} & \eqref{eq:CLU},\eqref{eq:DLU},  \eqref{eq:CompetiCons}. 
\end{array}
\eeq
The optimal value $\Pi(\overline{\Cbf},\underline{\Cbf})$ is DERA's profit given the distribution network access. In the objective function, DERA seeks profit maximization from both the aggregated customers' payments and the revenue from the wholesale market. Without loss of generality, all PoAs are under the same point of interconnection, facing a common LMP $\pi \in \mathbb{R}$. Optimal solutions and values in \eqref{eq:DERAsurplus_LnGPCC} support the control of aggregated DERs, offer/bid curves to the wholesale market ($\S$\ref{sec:Whosale}), and also the distribution network access request to DSO ($\S$\ref{sec:DERAinDSO}).

Under the above assumptions for feasibility in \eqref{eq:FeasibleAssump} and the utility function, Theorem~\ref {thm:DERA} below establishes a closed-form solution of \eqref{eq:DERAsurplus_LnGPCC} parameterized by $\pi$. We denote $V(x):=\frac{d}{dx} U(x)$ as  the marginal utility function, define \beq\label{eq:h}
h_{nk}(x):=\max \{\underline{d}_{nk},  \min \{V_{nk}^{-1}(x),\overline{d}_{nk}\}\}, 
\eeq 
and solve for $\underline{\xi}_m, \overline{\xi}_m, \forall m \in [M]$ from  
\begin{align} 
\sum_{n=1}^{{\cal N}_m}\sum_{k=1}^K  h_{nk}(\underline{\xi}_m) = \sum_{n=1}^{{\cal N}_m} g_n + \underline{C}_m,\label{eq:xiLB}\\
\sum_{n=1}^{{\cal N}_m}\sum_{k=1}^K  h_{nk}(\overline{\xi}_m) = \sum_{n=1}^{{\cal N}_m}g_n -\overline{C}_m.\label{eq:xiUB}
\end{align}
 
\begin{theorem}[{Optimal DERA scheduling and payment}] \label{thm:DERA} 
Given the wholesale market LMP $\pi$, the optimal consumption bundle $\dbf^*_n=(d^*_{nk})$ of prosumer $n$ and its  payment $\omega_n^*$ are given by
\begin{align}
d^*_{nk}&=\begin{cases}
    h_{nk}(\underline{\xi}_m), &  \sum\limits_{n=1}^{{\cal N}_m} g_n  \le \sum\limits_{n=1}^{{\cal N}_m}\sum\limits_{k=1}^K h_{nk}(\pi) -\underline{C}_m \\
    h_{nk}(\overline{\xi}_m ), &  \sum\limits_{n=1}^{{\cal N}_m} g_n \geq \sum\limits_{n=1}^{{\cal N}_m}\sum\limits_{k=1}^K h_{nk}(\pi)+\overline{C}_m \\
     h_{nk}(\pi), & \text{otherwise}
    \end{cases} \label{eq:d*}\\
\omega^{*}_n&= U_n(\dbf_n^*)-{\cal K}_n, \label{eq:omega*}
\end{align}
where $m$ is the index of PoA connecting customer $n$, \ie $n \in {\cal  N}_m$ defined by \eqref{eq:N}. 
\end{theorem}
The proof is in the appendix, following the convexity and  Karush-Kuhn-Tucker (KKT) conditions of (\ref{eq:DERAsurplus_LnGPCC}). This optimal solution has two noteworthy characteristics. First, the optimal consumption in \eqref{eq:d*} is only a function of the LMP $\pi$ when DERA purchases enough network accesses at the PoA. Note also the difference between the optimal consumption schedule in \eqref{eq:d*} and those in \cite{ Alshehri&etal:20TPS}, where the optimally scheduled consumption always depends on the anticipated LMP and forecast of BTM DG. Second, (\ref{eq:DERAsurplus_LnGPCC}) finds a Pareto efficient allocation that maximizes the surplus of the DERA, subject to the constraint that the aggregated customer has the given level of surplus ${\cal K}_n$. {The optimal payment $\omega_n^*$ is prosumer-specific. A similar optimization and the Pareto efficient allocation are also analyzed in \cite[P602]{Varian89PriceDiscri} for first-degree price discrimination. The payment function $\omega_n^*$ can be realized by a two-part tariff, which is explored in \cite{GaoAlshehriBirge:22},} although it unidirectionally aggregates BTM DG. Overall, such a closed-form solution allows the DERA to apply simple dispatch and pricing mechanisms when aggregating massive numbers of households over multiple PoAs in the distribution networks.

 \vspace{-0.3cm}
\subsection{NEM benchmarks} \label{sec:Benchmark}
Considering the benchmark performance of a regulated utility offering the NEM tariff, we extend the results in \cite{AlahmedTong2022NEM} and present closed-form characterizations of consumer/prosumer surpluses. For simplicity, we consider one representative prosumer by setting ${\cal N}_m=1, K=1$, and dropping the prosumer index $n$ and PoA index $m$. The prosumer's net consumption is $z := d-g,$ where $g\in [0, \infty)$ is the BTM DG.  The prosumer is a producer if $z<0$ and a consumer if $z\ge 0$.

In evaluating the benchmark prosumer surplus under a regulated utility, we assume that the prosumer maximizes its surplus under the utility's NEM tariff, where $\pi^+$ is the retail (consumption) rate, $\pi^-$ the sell (production) rate, and $\pi^0$ the connection charge. In general $\pi^-\le \pi^+$ under NEM tariff, and the prosumer's energy bill  $P(z)$ for the net consumption $z$ is given by the convex function $P(z) := \max\{\pi^+ z, \pi^-z\} + \pi^{\mbox{\tiny 0}}.$ The prosumer surplus under NEM is  $S(d):=U(d)-P(z).$

For an {\em active prosumer} whose consumption is a function of the available DG output $g$, the optimal consumption can be obtained by $
d_{\mbox{\tiny NEM-a}} := \arg\max_{d\in {\cal D}} (U(d)-P(d-g)).$ For the fairness of comparison,  we assume the aggregated customer is subject to the same distribution network injection and withdrawal access limits, \ie $-\underline{C} \le g-d \le \overline{C}$, which is the same as that applied to the proposed DERA optimization (\ref{eq:DERAsurplus_LnGPCC}). So, for the above optimization, the domain  is ${\cal D}:=[ \max\{\underline{d},g-\overline{C} \} , \min\{\bar{d},g+\underline{C}\}]$.

The surplus $S_{\mbox{\tiny NEM-a}}$  and the consumption $d_{\mbox{\tiny NEM-a}}$ of an active prosumer are given by the following equations.
\begin{align}\label{eq:acitiveSurplus}
    S_{\mbox{\tiny NEM-a}}(g, \underline{C}, \overline{C})&=U(d_{\mbox{\tiny NEM-a}})-P(d_{\mbox{\tiny NEM-a}}-g)\\ &=\begin{cases}
U(d^-)-\pi^-(d^--g)-\pi^0,& g\geq d^-\nn\\
U(d^+)-\pi^+(d^+-g)-\pi^0,&  g \le  d^+\nn\\
U(d^0)-\pi^0, &\text{otherwise}\\
\end{cases}\\
d_{\mbox{\tiny NEM-a}}&= \max \{d^+,\min \{g,d^-\}\}, \label{eq:dNEMa}
\end{align}
where $d^+:=f(\pi^+)$,  $d^-:=f(\pi^-)$, $d^0:=f(\mu^*(g))$ with 
\beq \label{eq:fconsump}
f(x):=\max \{\underline{d}, g-\overline{C}, \min \{V^{-1}(x),\bar{d}, g+\underline{C}\}\}, 
\eeq
and, by solving $f(\mu)=g$, we have $\mu^*(g)\in [\pi^-,\pi^+]$. 

A prosumer is called {\em passive} if it decides energy consumption without the awareness of its DG output and the influence brought by NEM X switching among $\pi^-$ and $\pi^+$. The optimal consumption bundle of such a  {\em passive prosumer} under the NEM X tariff is given by $
d_{\mbox{\tiny NEM-p}} :=  \arg\max_{d\in {\cal D}}  (U(d)-\pi^+d ).$ The total consumption $d_{\mbox{\tiny NEM-p}}  $ and the surplus $S_{\mbox{\tiny NEM-p}}$ of a  {\em passive prosumer} are given by
\begin{align}\label{eq:passiveSurplus}
    S_{\mbox{\tiny NEM-p}}(g, \underline{C}, \overline{C})&=U(d_{\mbox{\tiny NEM-p}})-P(d_{\mbox{\tiny NEM-p}}-g)\\ &=\begin{cases}
U(d^+)-\pi^-(d^+-g)-\pi^0,&g \geq d^+\nn\\
U(d^+)-\pi^+(d^+-g)-\pi^0,& g<d^+ 
\end{cases}\\
d_{\mbox{\tiny NEM-p}}&=  d^+.\label{eq:dNEMp}
\end{align}
In summary, the prosumer surplus under NEM, $S_{\mbox{\tiny NEM}}(g, \underline{C}, \overline{C})$ is given by
\begin{align}\label{eq:NEMS}
S_{\mbox{\tiny NEM}}(g, \underline{C}, \overline{C})=\begin{cases}
S_{\mbox{\tiny NEM-a}}(g, \underline{C}, \overline{C}),& \text{active prosumer},\\
S_{\mbox{\tiny NEM-p}}(g, \underline{C}, \overline{C}),& \text{passive prosumer}. 
\end{cases}
\end{align}

\subsection{Properties of DERA Competitive with NEM}\label{sec:CNEM}

We analyze the profitability of DERA and the energy consumption cost of aggregated prosumers when DERA is competitive with a regulated NEM tariff parameterized. Assume $0 \le \pi^-\le \pi^+$ and customers' surpluses under NEM are nonnegative \cite{AlahmedTong2022NEM}. From \eqref{eq:NEMS}, we have the $n$-th prosumer surplus under NEM $S_n^{\mbox{\tiny NEM}}(g_n, \underline{C}_n, \overline{C}_n)$, whose computation depends on the DG generation and network access limits. DERA sets the benchmark prosumer surplus  
\beq\label{eq:NEMK}
{\cal K}_n  = \zeta S_n^{\mbox{\tiny NEM}}(g_n, \underline{C}_n, \overline{C}_n), ~~\zeta \ge 1,
\eeq
which is used in \eqref{eq:CompetiCons} with profit ratio $\zeta $ to obtain competitive aggregation over the DSO's NEM-based aggregation with the same network access.\footnote{Customers owning DERs switch from NEM to DERA for higher consumer surplus, granting DERs control to DERA upon joining.} In this subsection, the network access limits carry the subscript $n$, which is equivalent to $m$ since we set ${\cal N}_m = 1$ and $K = 1$ for simplicity.

The {\em ${\cal K}$-competitive constraint} in \eqref{eq:CompetiCons} has significant implications on pricing stability, despite the fact that the aggregation is based on real-time LMP. Price stability means the price and payment faced by customers cannot go randomly high, for which a counterexample is the real-time LMP. Because the NEM tariff has price stability, achieving a finite customer payment regardless of the wholesale LMP fluctuation, an aggregation mechanism competitive with the NEM tariff must also be stable. The proposition below formalizes this intuition. 
\begin{proposition}[Average cost of consumption]\label{prop:Pcappi+} The prosumer’s average energy consumption cost under DERA aggregation is no higher than the NEM retail rate,  \ie $\omega_n^*/d_n^* \le \pi^+$.
 \end{proposition} 
\noindent See proof in the appendix. Such price stability comes directly from the {\em ${\cal K}$-competitive constraint}, which enforces a lower bound for customer surplus and thus naturally limits the maximum customer payment. Note that the two-part pricing of Griddy \cite{Griddy} is not a stable pricing mechanism because the retail rate is tied directly to real-time LMP. 

In the {\em ${\cal K}$-competitive constraint} \eqref{eq:CompetiCons} with \eqref{eq:NEMK}, the profit ratio $\zeta$ controls surplus distribution between  DERA and its aggregated prosumers. A larger $\zeta$ rebates more benefits to prosumers and incentivizes prosumers to join DERA, although it increases the deficit risk of DERA. Therefore, the DERA must carefully set $\zeta$ to balance profitability and competitiveness. In Proposition~\ref{prop:RA}, we establish that the DERA can achieve nonnegative expected profit by choosing an appropriate $\zeta$, assuming that BTM DG generation $\gbf$ and the LMP $\pi$ are independent random variables in a competitive market. 
\begin{proposition}[{Profitability of DERA}] \label{prop:RA} If $\pi^-\le \mathbb{E}[\pi]\le \pi^+$, then there exists a profit ratio $\zeta\geq 1$ such that the  DERA's expected profit is nonnegative. 
\end{proposition}
The proof is provided in the appendix. In practice, the condition $\pi^- \le \mathbb{E}[\pi] \le \pi^+$ is often satisfied. For instance, in many states—including California—the export rate $\pi^-$ is set near the avoided cost, which typically approximates the expected LMP $\mathbb{E}[\pi]$, as a way to mitigate cross-subsidies \cite{borenstein2021designing}.

{During LMP spikes, we assume the regulated retail consumption rate $\pi^+$ remains capped and is not influenced by price spikes. Consequently, the average energy consumption cost from Proposition~\ref{prop:Pcappi+} is bounded by $\pi^+$ and insensitive to these spikes. While customers/prosumers are shielded from wholesale LMP spikes, both the incumbent utility and DERAs may experience negative profits during these events. Because LMP spikes are infrequent, the DERA’s average surplus remains positive (Proposition~\ref{prop:RA}).}

\section{DERA Wholesale Market Participation}\label{sec:Whosale}

{The virtual storage model is adopted by ISOs to enable DERAs' participation in the wholesale market with bi-directional monetary and power flows \cite{ISONE19CSF, ISONE21}. This means DERA can submit a combination of supply offers and demand bids, purchasing its aggregated consumption (as charging the virtual storage) and selling its aggregated production (as discharging). The state-of-charge (SOC) dynamics and multi-period formulations are not considered in this paper, as SOC models vary across ISOs and not all real-time electricity markets implement multi-period optimization.}
 
 \vspace{-0.3cm}
\subsection{Offer/Bid Curves of DERA in Energy Markets}
As a virtual storage participant in the real-time energy market, the DERA is either self-scheduled or scheduled by ISO/RTO according to its bids and offers. This work focuses on developing price-quantity bid/offer curves that define DERA's willingness to consume/produce. In a competitive market, such curves are the marginal cost of production and the marginal benefit of consumption derived from the optimal DERA decision in Theorem~\ref{thm:DERA}.

Let $Q$ be DERA's aggregated quantity to buy (when $Q<0$) or sell (when $Q >0$), and $\pi$ be the wholesale market LMP.  Let $G:=\sum_{n=1}^N g_n$  be the BTM DG aggregated by DERA. In a competitive market, a price-taking DERA participant bids truthfully with its aggregated supply function 
\beq\label{Q-P}
\begin{array}{l}
Q=\Fmsc(\pi), ~~\Fmsc(\pi ) := G - \sum_{n=1}^N\mathbf{1}^\top \dbf^*_n(\pi ),
\end{array}
\eeq
where $\dbf^*_n$ is defined in \eqref{eq:d*}. Note that the inverse of the DERA supply function  $\Fmsc^{-1}(Q)$ defines the offer/bid curves of the DERA. For a quantity bid, the DERA forecasts the LMP and computes the optimal net production with \eqref{Q-P}. In contrast, for a price-quantity bid/offer curve $\Fmsc^{-1}(Q)$, the DERA avoids LMP forecasting, as the ISO clears the market using the submitted curve and ensures consistency between the LMP and the resulting dispatch. More details are provided in Lemma~\ref{lemma:WMC}, and a simulation of this offer/bid curve is presented in our previous paper \cite{chen22competitive}. 

Note also that the supply function depends on the aggregated BTM generation $G$, which is not known to the DERA at the time of the market auction. In practice, $G$ can be approximated by using historical data or $ N \mbbE(g_n)$ via the Law of Large Numbers involving $N$ independent prosumers or via the Central Limit Theorem for independent and dependent random variables \cite{Billingsley:95}. 

\vspace{-0.3cm}
\subsection{Market Efficiency with DERA Participation}

{We now establish that the DERA's participation in the wholesale market achieves the same social welfare as that when all profit-maximizing prosumers participate in the wholesale market autonomously.  We assume the wholesale market is competitive, where all participants are price takers with truthful bidding incentives. The theoretical result here is parallel to traditional competitive equilibrium analysis in the electricity market.} Prosumer notations in this section overlap with those in $\S$\ref{sec:Retail}, but subscripts are modified to include the transmission network bus index.

Consider a transmission network with $I$ buses. At each bus of the transmission network, we assume $M$ PoAs at the distribution network are connected, and $N$ prosumers are aggregated by the proposed DERA model. Denote $U_{in}$ as the concave utility function for the $n$-th prosumer at the $i$-th transmission network bus. $\gbf_n:=(g_{in})_{i\in[I]}$ and $\dbf_n:=(d_{in})_{i\in[I]}$  are respectively the vectors of BTM DG generation and energy consumption for the prosumers. For simplicity, we ignore the number of energy-consuming devices for each prosumer, \ie $K=1$, in this section. At each transmission bus, we sum up all load-serving entities and generators into one demand function and supply function. The load-serving entity at bus $i$ purchases electricity $e_i$ with a concave benefit function $B_i(e_i)$. The generator at bus $i$ produces $p_i$ with a convex cost function $C_i(p_i)$. Denote $\ebf:=(e_i)_{i\in [I]}$, $\pbf:=(p_i)_{i\in [I]}$. $ \fbf \in \mathbb{R}^L$ is the line flow limit for $L$ branches of the transmission network. $ \Sbf\in \mathbb{R}^{L\times I}$ is the network parameter for DC power flow model.    
\begin{lemma}[Wholesale market clearing with DERA] \label{lemma:WMC}  
When prosumers participate in the wholesale market indirectly through the proposed DERA with offer/bid curve (\ref{Q-P}), social welfare $\mbox{\sf SW}_{\mbox{\rm\tiny DERA}}$ is the optimal value of the convex problem
\vspace{-0.2cm}
\begin{subequations}\label{eq: SS_CC}
\begin{align}
&\underset{\Dbf, \pbf, \ebf\geq 0}{\rm max}&&~~ \sum_{i=1}^I  (\sum_{n=1}^N U_{in}(d_{in})+B_i(e_i)- C_i(p_i))\label{eq:SWDERAobj}\\
&   \text{subject to}  && \eqref{eq:CLU},\eqref{eq:DLU},\nn\\
&\lambda:& &\sum_{i=1}^I p_i=\sum_{i=1}^I(\sum_{n=1}^N (d_{in}-g_{in})+e_i),\label{eq: PB}\\
&\mubf:&& \Sbf (\sum_{n=1}^N (\gbf_n-\dbf_n)+\pbf-\ebf) \preceq \fbf.\label{eq: network}
\end{align}
\end{subequations}
The sum of the DERA surplus and prosumers' surpluses, denoted by $ S_{\mbox{\tiny DERA}}$, can be computed by
\beq\label{eq:SWDERA}
\begin{array}{l}
S_{\mbox{\tiny DERA}}=\sum_{i=1}^I \sum_{n=1}^N (U_{in}(d^{\star}_{in})-\pi_i (d^{\star}_{in}-g_{in})), 
\end{array}
\eeq
where $d^{\star}_{in}$ is the optimal solution of \eqref{eq: SS_CC}, which equals \eqref{eq:d*}.
\end{lemma}
Proof of this Lemma in our appendix relies on showing that pricing and dispatch results from \eqref{eq: SS_CC} are at the bidding curve of DERA, \ie \eqref{Q-P}. With the optimal dual  $\lambda^{\star}\in \mathbb{R}$ for the power balance constraint \eqref{eq: PB} and $\mubf^{\star}\in \mathbb{R}^L$ for the line flow limit \eqref{eq: network}, the market clearing LMP over $I$ buses is defined by  $\pibf : =\mathbf{1}\lambda^{\star}-\Sbf^\top \mubf^{\star}$, 
where $\pibf :=(\pi_i)_{i\in [I]}$. The prosumer utility in \eqref{eq:SWDERAobj}, and constraints for energy consumption and distribution network access in \eqref{eq:CLU}\eqref{eq:DLU} come from DERA's offer/bid (\ref{Q-P}).  


As for the prosumer's direct participation in the wholesale market, a price-taking prosumer $n$ and bus $i$  constructs her offer/bid curves by solving the following surplus maximization problem with the given LMP $\pi_i$:
\beq\label{eq:ProfitMaxDirect}
\begin{array}{l}
\underset{d_{in} \in  {\cal D}_{in}}{\rm max}~~ U_{in}(d_{in})-\pi_i (d_{in}-g_{in}),\end{array}\eeq
where ${\cal D}_{in}:=[\max\{\underline{d}_{in},g_{in}-\overline{c}_{in} \}, \min\{\bar{d}_{in},g_{in}+\underline{c}_{in}\}]$. The access limits $\overline{c}_{in}$ and $\underline{c}_{in}$ represent the distribution network injection and withdrawal capacities allocated to each prosumer. Detail formulations for $\overline{c}_{in}$ and $\underline{c}_{in}$ are in \eqref{eq:AccessPropsumerUB}\eqref{eq:AccessPropsumerLB} of our appendix. These values are consistent with \eqref{eq:d*}, ensuring a fair comparison. Under this setup, prosumers participating directly in the wholesale market face the same access constraints at each PoA as those in our proposed DERA model. Solve \eqref{eq:ProfitMaxDirect} and obtain the bid/offer curve for prosumer $n$ at bus $i$, \ie 
\beq
\begin{array}{l}\label{Q-P_prosumer}
 \Smsc_{in}(\pi_i)= g_{in}-d^*_{in}(\pi_i),
\end{array}
\eeq
where $d^*_{in}(\pi_i)$ takes the same definition as that in \eqref{eq:d*}.  

Let $\mbox{\sf SW}_{\mbox{\rm\tiny Direct}}$ and $S_{\mbox{\rm\tiny PRO}}$ be, respectively, the optimal social welfare and prosumers' surplus when all prosumers directly participate in the wholesale market. {The following theorem parallels the market efficiency result in \cite{GaoAlshehriBirge:22} and traditional competitive equilibrium derivations, although we consider different aggregation methods and  distribution network access. }
 \vspace{-0.3cm}
\begin{theorem}[{Market efficiency}] \label{thm:MarketEfficiency}  
When all prosumers directly participate in the wholesale market, the market-clearing result can be computed by \eqref{eq: SS_CC}, $\mbox{\sf SW}_{\mbox{\rm\tiny Direct}}=\mbox{\sf SW}_{\mbox{\rm\tiny DERA}}$, and $S_{\mbox{\rm\tiny PRO}}=S_{\mbox{\rm\tiny DERA}}.$
\end{theorem}

\noindent The proof is provided in the appendix, which relies on the fact that the proposed DERA has its bidding curve \eqref{Q-P}  equal to the sum of the prosumer's bidding curve in \eqref{Q-P_prosumer}. From this, we can establish that the wholesale market clearing problem with the direct participation of all prosumers has the same market-clearing results as  (\ref{eq: SS_CC}). 

Although the proposed DER aggregation model only focuses on DERA's profit maximization in the objective of \eqref{eq:DERAsurplus_LnGPCC}, the competitive constraint \eqref{eq:CompetiCons} aligns the aggregated prosumer's surplus maximization with DERA's profit maximization. So, the proposed competitive DER aggregation has the incentive to maximize prosumers' surpluses and get the maximum total surplus that can be split among DERA and its aggregated prosumers. Essentially, the DERA acts as an intermediary, enabling prosumers to indirectly participate in the wholesale market. As the DERA earns a profit for providing this service, each prosumer receives a lower surplus than they would under direct participation (illustrated in Fig.~\ref{fig:SurplusDS}). This is justified, since individual prosumers lack the scale required for direct participation in the wholesale market.

\section{DERA-DSO coordination}\label{sec:DERAinDSO}

All generation and consumption resources aggregated by DERA need to bypass the distribution network to participate in the wholesale market. The DERA aggregation presented in this work ensures that the aggregated DER at each distribution network PoA is bounded by access limits imposed through the distribution network access limit auction in \cite{ChenBoseMountTong23DERA}. A DERA submits a bid curve in this auction representing its willingness to acquire access at PoAs. We assume that a DERA is a price taker in the access limit auction.  Therefore, the bid-in demand curve for network access from the DERA at a particular PoA is the marginal benefit (profit) from having a DER aggregation under the PoA. The maximum expected profit of DERA is   
\beq \varphi(\overline{\Cbf},\underline{\Cbf}) := \mbbE_{\gbf, \pi}[\Pi(\overline{\Cbf},\underline{\Cbf})],
\eeq
where $\Pi(\overline{\Cbf},\underline{\Cbf})$ is the maximum DERA profit computed from the optimal value of (\ref{eq:DERAsurplus_LnGPCC}), given the realized renewable generations over all buses and the realized LMP. Note that when participating in the forward network access auction, both the BTM DG and LMP are random.
 \vspace{-0.3cm}
\subsection{DERA Benefit Function for Distribution Network Access}
The following Proposition provides an expression for the benefit function of DERA, $\varphi(\overline{\Cbf},\underline{\Cbf})$, which can be used as the bid curve of access limits submitted to the auction in \cite{ChenBoseMountTong23DERA}.
\begin{proposition}[Benefit function for network access]\label{Prop:DERA} With the DERA profit maximization (\ref{eq:DERAsurplus_LnGPCC}),
the expected DERA surplus is 
\begin{align} 
&\varphi({\underline{\Cbf}},{\overline{\Cbf}})=\mbbE\big\{ \sum_{m=1}^M (\underline{\phi}_m({\underline{C}_m})+ \overline{\phi}_m({\overline{C}_m}))+ \sum_{n=1}^N (\varrho_n -{\cal K}_n)  \big\},\label{eq:bidCurveall}\\
&\underline{\phi}_m({\underline{C}_m}):= \big(\sum_{n=1}^{{\cal N}_m} U_{n}(\hbf_n(\underline{\xi}_m))-\pi \underline{C}_m\big)\mathbbm{1}\{ \sum_{n=1}^{{\cal N}_m} g_n \le   \underline{q}_m\},\nn
\\
&\overline{\phi}_m({\overline{C}_m}):=\big(\sum_{n=1}^{{\cal N}_m}U_{n}(\hbf_n(\overline{\xi}_m))+\pi \overline{C}_m\big)\mathbbm{1}\{\overline{q}_m \le \sum_{n=1}^{{\cal N}_m} g_n \},\nn
\\
& \varrho_n:= \big(U_{n}(\hbf_n(\pi))-\pi(\hbf_n(\pi)-g_n)\big)\mathbbm{1}\{   \sum_{n=1}^{{\cal N}_m} g_n  \in (\underline{q}_m, \overline{q}_m)\}.\nn
\end{align}
\end{proposition}
\noindent The proof is provided in appendix with $ \hbf_n(x) := \sum_{k=1}^K h_{nk}(x)$, $U_{n}(\hbf_n(x)) = \sum_{k=1}^K U_{nk}(h_{nk}(x))$ from the additive property of the utility function, and 
\beq
\begin{array}{l}
\overline{q}_m :=\overline{C}_m +\sum_{n=1}^{{\cal N}_m}\hbf_n(\pi),
~\underline{q}_m :=-\underline{C}_m+\sum_{n=1}^{{\cal N}_m}\hbf_n(\pi).\nn
\end{array}
\eeq 
The optimal DERA surplus is decomposed into three terms: one dependent on the withdrawal access $\underline{\phi}_m({\underline{C}_m})$, one dependent on the injection access $\overline{\phi}_m({\overline{C}})$, and one independent of network access. $\varphi({\underline{\Cbf}},{\overline{\Cbf}})$ is separable across injection and withdrawal access over $M$ PoAs. Therefore, DERA can bid separately for the distribution network accesses at different PoAs when coordinating with DSO. At PoA $m$ with less BTM DG, \ie $ \sum_{n=1}^{{\cal N}_m} g_n \le   \underline{q}_m$, DERA's benefit depends on the withdrawal access. Conversely, if there is more  BTM DG, \ie $\overline{q}_m \le \sum_{n=1}^{{\cal N}_m} g_n$, DERA's benefit depends on the injection access. Related simulations are in $\S$\ref{sec:accessEQ}. 
  
 \vspace{-0.2cm}
\subsection{Long-Run Equilibrium for Competitive DERA}

In a long-run competitive industry, we explore how many DERA can survive. DERAs compete to attract customers, attain distribution network access, and participate in the wholesale market. We assume all DERAs adopt the competitive DER aggregation method in \eqref{eq:DERAsurplus_LnGPCC}. The condition for a competitive long-run equilibrium \cite[P193]{jehle2011advanced} has two components: (i) the marginal benefit of DERA equals the marginal cost of DSO for providing the distribution network access, and (ii) all DERAs have profits equal to zero, \ie DERA's profit in conducting aggregation equals DERA's payment to acquire distribution network access. Related derivations and simulations are in $\S$\ref{sec:EQ_MultiT} and  our  appendix.

\section{Case Studies}\label{sec:CaseStudies}

We compared the expected surplus distribution of different DER aggregation methods under varying network access limits and BTM DG generations.\footnote{Under the access limit allocation framework in \cite{ChenBoseMountTong23DERA}, distribution network reliability concerns are resolved if DERAs obey allocated distribution network access limits. So the distribution network topology was ignored here.} We also computed the benefit function of DERA to the distribution network access and empirically evaluated the long-run equilibrium of DERA with multi-interval aggregation. 
  \vspace{-0.2cm}
\subsection{Parameter Settings}\label{sec:param}
Denote the utility function for the aggregated customer as
\beq \label{eq:UtilityF}
U(x)=\begin{cases}
\alpha x-\frac{\beta}{2}x^2,&0\le x\le \frac{\alpha}{\beta}\\ 
\frac{\alpha^2}{2\beta},&x> \frac{\alpha}{\beta}
\end{cases},
\eeq 
where $\alpha=\$0.4/\mbox{kWh}, \beta=\$0.1/(\mbox{kWh})^2$ \cite{AlahmedTong2022NEM}. Let the marginal utility $ V^{-1} \in [\underline{d}, \bar{d}]$ for the consumption limits.\footnote{We simulated the case of $K=1$ for simplicity.} 

We use NEMa and NEMp to represent the DER aggregation under NEM when prosumers were active and passive, respectively. Passive customers are not responsive to the retail prices, but active customers will optimize their energy consumption given the retail price and the BTM DG generations. Based on PG\&E residential rate, we set  $\pi^+=\$0.3/\mbox{kWh}$ for the NEM. We assumed $\pi^{-}=\mathbb{E}[\pi]$ and the fixed cost of NEM was covered by extracting a fixed payment from DERA, so we simulated with $\pi^0=\$0$. Gao-Alshehri-Birge (GAB) represented the two-part pricing in \cite{GaoAlshehriBirge:22}, which allowed customers to sell BTM DG to the DERA while purchasing energy from its incumbent utility company. Detailed models for NEMa, NEMp, and GAB are provided in Appendix $\S$\ref{sec:Benchmark} and the appendix. Our DER aggregation method was simulated in Co.NEMa and Co.GAB, competitive with NEMa and GAB, respectively. For Co.NEMa, we set the profit ratio $\zeta$ at the upper bound in the proof of Proposition~\ref{prop:RA}. For Co.GAB, we set $\zeta=1.05$ to provide 5\% more customer surplus than the GAB competitor. 

We considered the randomness of LMP and BTM DG generation using data sources from CAISO \cite{CAISO25price} and Pecan Street Dataport \cite{PECAN}, respectively. The LMP $\pi$ was modeled as a Gaussian random variable\footnote{{The mean of LMP primarily drives expected surplus of both customers and DERAs, while LMP distribution and STD have minor effects. To simulate price spikes and increased variability, we conducted additional simulations by modeling LMP as (i) a lognormal variable with scale  \$0.05/kWh and shape 0.55, and (ii) a Gaussian variable with mean  \$0.05/kWh and STD \$0.05/kWh.  Both model results align with Fig.~\ref{fig:DS}.}}with a mean of \$0.05/kWh and a standard deviation (STD) of \$0.01/kWh. The BTM DG generation $g$ was modeled as a Gaussian random variable with a mean ranging from 1.1 kWh to 5.1 kWh and a standard deviation of 0.2kWh, truncated at $(0,+\infty)$. We generated 10,000 random scenarios for both the LMP and BTM DG. At a given PoA, we evaluated the expected per-customer surplus metrics based on the sample means from these scenarios.

 \vspace{-0.2cm}
\subsection{Performances with Unlimited Distribution Network Access}\label{sec:4Cases}

Four observations below were drawn when all aggregators received plenty of distribution network accesses. 

\begin{figure}
\centering
\scalebox{0.42}{\includegraphics{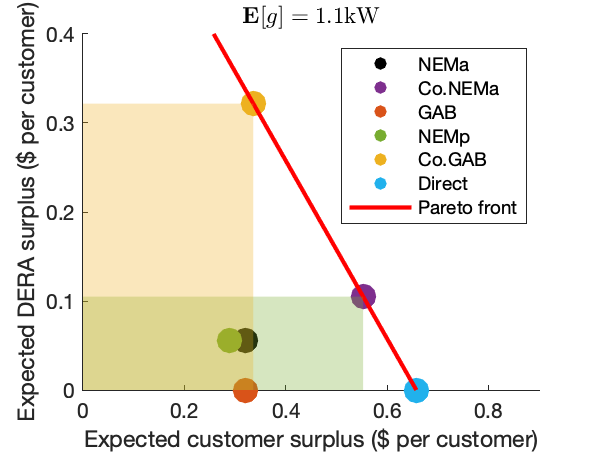}} \scalebox{0.42}{\includegraphics{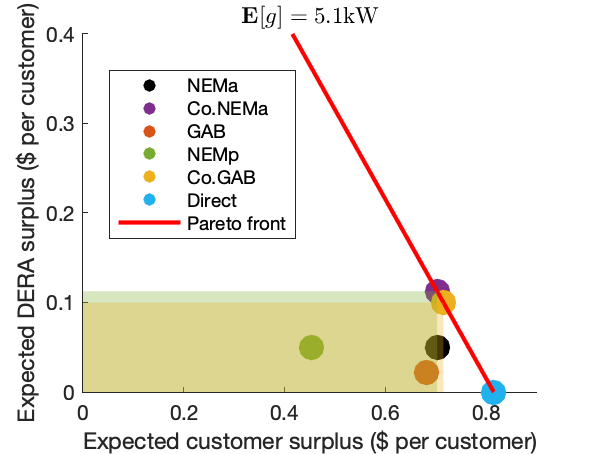}} 
 \vspace{-0.2cm}
\caption{Expected surplus distribution and market efficiency with 80\% DG adopter rate. Each shaded rectangle is dominated by its top right corner. (From left to right, the expected DG increases from 1.1 to 5.1 kW.)}
 \vspace{-0.3cm}
\label{fig:SurplusDS}
\end{figure}

First, Co.NEMa and Co.GAB were at the Pareto front in Fig.~\ref{fig:SurplusDS}, achieving the maximum social surplus as if all prosumers directly participated in the wholesale market. This verified Theorem~\ref{thm:MarketEfficiency}. The Pareto front was computed by aggregating the surpluses of DERA and customers, omitting the surpluses of other units. This was because we adopted the price taker assumption in the wholesale market, thus surpluses of other units stayed the same in different DERA models. The blue dot, named Direct, represented the ideal case that prosumers directly participated in the wholesale market with bidding curve \eqref{Q-P_prosumer}. The green rectangle contained aggregation methods achieving less DERA surplus and customer surplus than Co.NEMa, thus dominated by our proposed competitive DER aggregation method. Similarly, the orange rectangle was dominated by its top right corner, Co.GAB. This was because our aggregation methods efficiently participated in the wholesale market with aggregated resources and scheduled the aggregated customers at a consumption level with a higher customer surplus. When the expected BTM DG increased from 1.1kW to 5.1kW, comparing the left and right panels in Fig.~\ref{fig:SurplusDS}, we observed that the expected social surplus, which was the sum of DERA and customer surpluses, increased, because more BTM DG was sold to the wholesale market. 

\begin{figure}
\centering
 \vspace{-0.3cm}
  \scalebox{0.43}{\includegraphics{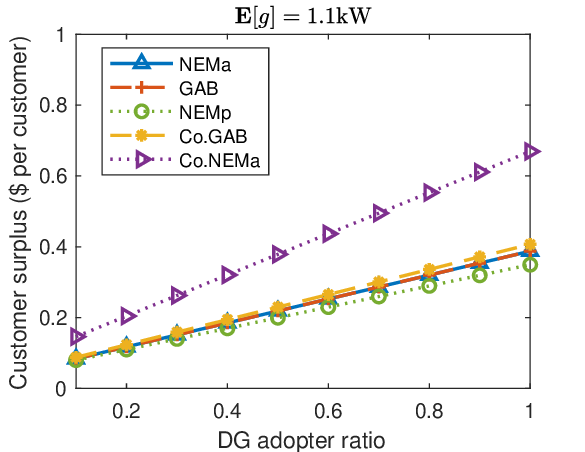}}  \scalebox{0.43}{\includegraphics{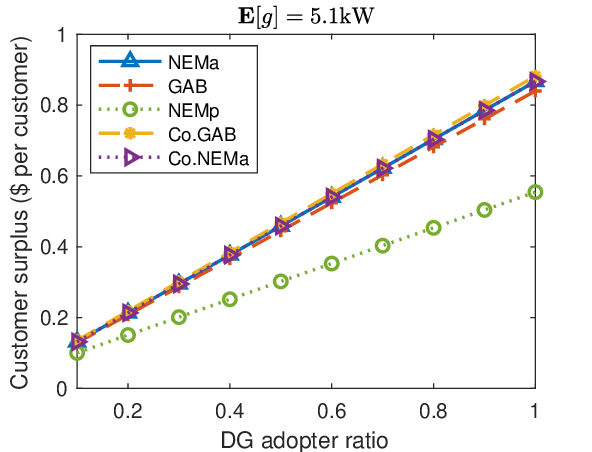}}
 \scalebox{0.44}{\includegraphics{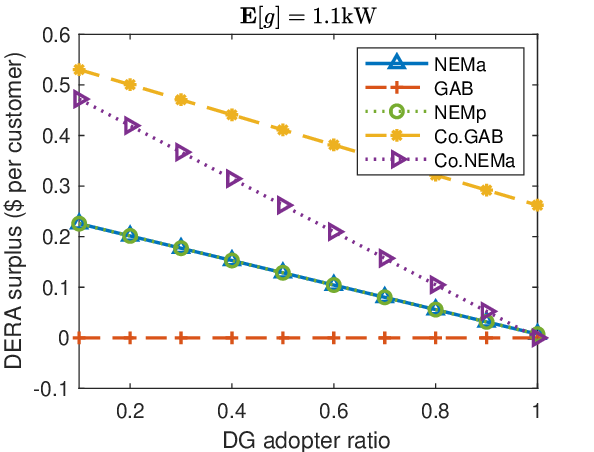}}\scalebox{0.44}{\includegraphics{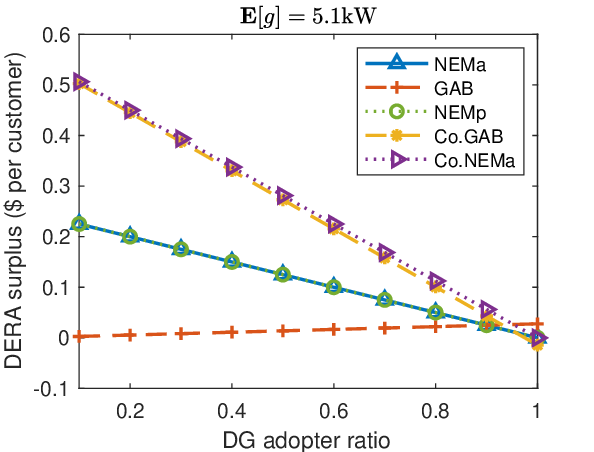}}
 \vspace{-0.5cm}
\caption{Expected surplus distributions vs. network access ratio. (Top: expected customer surplus; bottom: expected DERA surplus.) }
 \vspace{-0.3cm}
\label{fig:DS}
\end{figure}

Second, customers had the highest expected surplus in Co.NEMa and Co.GAB, as shown by the top of Fig.~\ref{fig:DS}. Passive customers in NEMp had the least surplus because their scheduling was agnostic of DG generation. Customer surpluses almost overlapped in all cases at a low DG adopter ratio with fewer producers, since most aggregation benefits came from BTM DG of producers. When the DG adopter ratio increased, the expected customer surplus increased in all cases. 

Third, when the DG adopter ratio or the DG generation was low, Co.NEMa and Co.GAB achieved the highest expected DERA surplus, as shown at the bottom of Fig.~\ref{fig:DS}. When the DG adopter ratio and DG generation were high, GAB achieved the highest DERA surplus because GAB only aggregated producers.\footnote{GAB achieved the Pareto front when all prosumers were producers, e.g., DG adopter ratio equal 100\% and $\mathbb{E}[g]=5.1$kW.} Co.NEMa always had DERA profit no less than zero since we chose $\zeta$ based on Proposition~\ref{prop:RA}. The DERA surpluses under NEMp and NEMa were identical, as setting $\pi^-=\pi$ ensures that aggregated DGs are compensated at the wholesale market price, eliminating any surplus gained by aggregating active prosumers' DG production.

Fourth, since NEM provided a higher surplus to customers with BTM DG, DERAs must commensurately reduce their profits and share them with the customers to remain competitive with NEM. Therefore, in most cases of Fig.~\ref{fig:DS}, the expected DERA surplus decreased when the DG adopter ratio increased. However, GAB had an increasing DERA surplus when the DG adopter ratio increased since GAB only aggregated producers. 

 \vspace{-0.2cm}
\subsection{Performances with Limited Distribution Network Access} 
\begin{figure}
\centering
 \vspace{-0.3cm}
 \scalebox{0.44}{\includegraphics{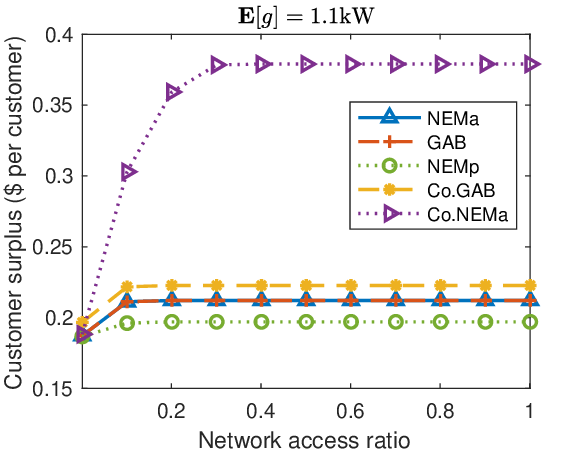}}\scalebox{0.44}{\includegraphics{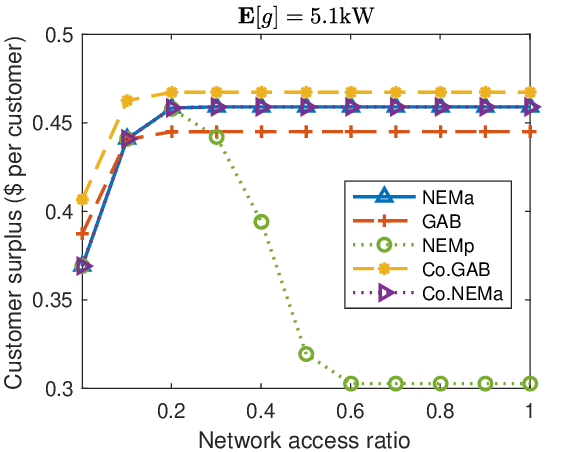}}
\scalebox{0.44}{\includegraphics{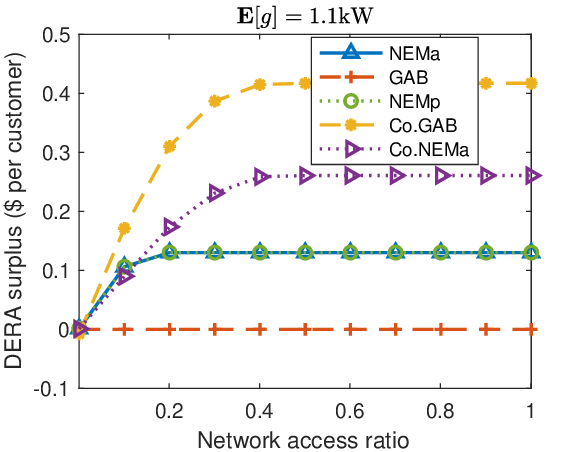}}\scalebox{0.44}{\includegraphics{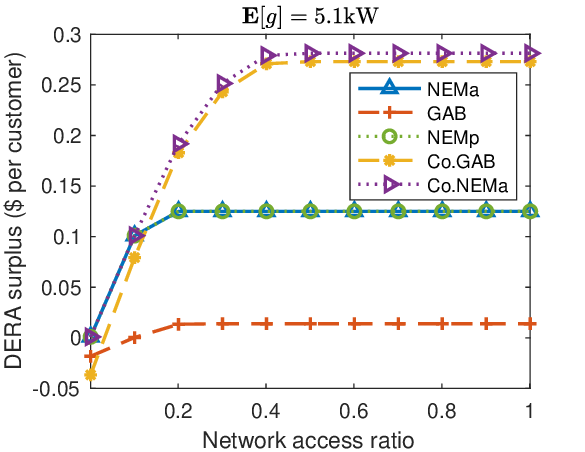}}
 \vspace{-0.2cm}
\caption{Expected surplus distributions vs. network access ratio with 50\% DG adopter rate. (Top: expected customer surplus; bottom: expected DERA surplus when $\mathbb{E}[g]=$ 1.1 kW and 5.1 kW, respectively.) }
 \vspace{-0.6cm}
\label{fig:LimitAccess}
\end{figure}

Here, we set distribution network access limits for each prosumer by $\overline{C}=\underline{C}= 8\delta$ kW and varied the {\em network access ratio} $\delta$ from 0 to 1 to analyze the influence of limited distribution network access. First, as is shown in Fig.~\ref{fig:LimitAccess}, either Co.NEMa or Co.GAB achieved the highest customer surplus or DERA surplus under a limited network access ratio. Second, when the network access ratio increased, customer surplus increased in most cases except NEMp, which passively controlled DG. Third, the DERA surplus in all cases increased when the network access ratio increased. This was intuitive because DERAs needed distribution network access to deliver the aggregated DER and participate in the wholesale market. 

\vspace{-0.2cm}
\subsection{Benefit Function of DERA for Distribution Network Access }\label{sec:accessEQ}

\begin{figure}[htbp]
    \centering
     \vspace{-0.3cm}
    \includegraphics[scale=0.62]{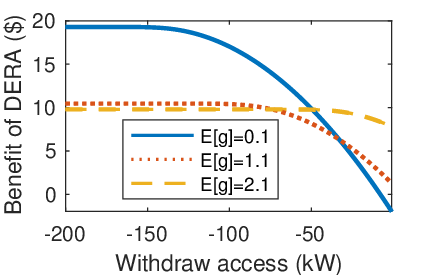}\includegraphics[scale=0.62]{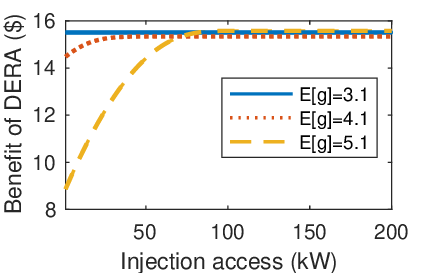}  
     \vspace{-0.3cm}
    \caption{DERA Benefit function $\varphi$. (Left:  withdrawal access $-\underline{C}$; right: injection access $ \overline{C}$.)}
     \vspace{-0.5cm}
    \label{fig:INJ}
\end{figure}

We computed the bid-in benefit function of the proposed DERA model, \ie \eqref{eq:bidCurveall}, with $\zeta=1.01$ and 50 prosumers aggregated at a certain PoA. DERA was competing with NEM, and prosumers were passive. Figure~\ref{fig:INJ} shows the expected benefit $\varphi$ of the DERA as a function of injection and withdrawal access, under varying levels of expected BTM DG generations.

In Fig.\ref{fig:INJ} (left), DERAs with lower expected DG generation exhibited higher benefits and submitted higher bid prices for withdrawal access, as indicated by the steeper slope of the benefit function. This is because, with less BTM generation, DERAs rely more on electricity withdrawn from the network. In Fig.\ref{fig:INJ} (right), the benefit function decreased with higher DG generation—a counterintuitive result. This occurred because NEM offers greater surplus to customers with higher DG output, forcing DERAs to reduce their profit margins and share more benefits with customers to remain competitive.

\vspace{-0.2cm}
\subsection{Long-Run Competitive Equilibrium of DERAs}\label{sec:EQ_MultiT}

In the long-run competitive equilibrium analysis with multi-interval aggregation of DERs, we assumed that 200 DERAs initially existed and computed the expected number of surviving DERAs in the long run. For simplicity, we assumed DERAs were homogeneous and had the same setting as $\S$\ref{sec:accessEQ}. Prosumers had the same expected DG generation created from the 24-hour rooftop solar data in Pecan Street.\footnote{Detail DG trajectories and the long-run equilibrium results for single-interval aggregation are shown in the   appendix, providing intuitions about long-run equilibrium for multi-interval aggregation here.} We multiplied the mean of 24-hour DG by the factor $\epsilon_1 \in \mathbb{R}_+$ to simulate different DG installation capacities and sampled 10,000 random DG scenarios. DERA submitted the benefit function, as in Fig.~\ref{fig:INJ}, to acquire hourly distribution network access. Same as \cite{ChenBoseMountTong23DERA}, the DSO cost function for providing distribution network access was assumed to be the sum of quadratics, $J(x)=\frac{1}{2}bx^2+ax$ with  $a=\$0.009/\mbox{kWh}, b=\$0.0005/(\mbox{kWh})^2$ for both the injection and withdrawal access. We multiplied DSO's cost $J$ by $\epsilon_2\in \mathbb{R} _+$ to simulate different levels of DSO's costs.

\begin{figure}[htbp]
    \centering
     \vspace{-0.3cm}  
    \includegraphics[scale=0.47]{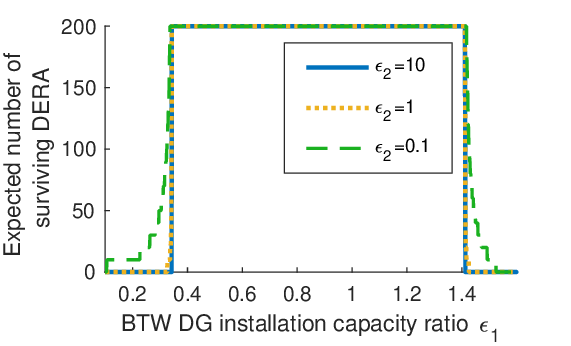}\includegraphics[scale=0.47]{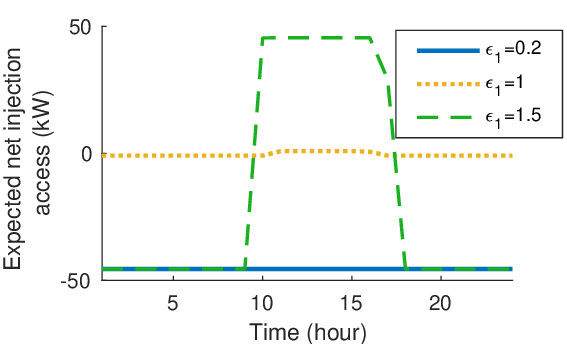}
     \vspace{-0.8cm}
    \caption{Long-run competitive equilibrium for multi-interval aggregation. (Left: expected number of surviving DERA vs. DG installation capacity ratio $\epsilon_1 $; right: expected distribution network net injection access of DERA over 24 hours, whose negativity represents withdrawal access.)}
     \vspace{-0.4cm}
    \label{fig:LREQ}
\end{figure}

Two observations were drawn from the results in Fig.~\ref{fig:LREQ}. First, when the DG capacity ratio $\epsilon_1$ was between 0.4 and 1.4, all initial 200 DERAs survived because DERAs were able to internally balance customer demands with their aggregated DG, thus relying less on and paying less to the network access. This was validated by the yellow dot curve from Fig.~\ref{fig:LREQ} (right), which required almost zero network access over 24 hours. Second, when the DG capacity ratio decreased from 0.4 to 0 in Fig.~\ref{fig:LREQ} (left), the number of surviving DERA decreased. In this case, DG was lower than the aggregated customers' consumption, and not all DERAs can survive when competing and paying for network withdrawal access over 24 hours, as shown by the blue solid curve in Fig.~\ref{fig:LREQ} (right). In the green dashed curve of Fig.~\ref{fig:LREQ} (left), DSO's cost for providing network access was lower, so more DERAs survived than in other curves. Similar reasons applied when the DG capacity ratio increased beyond 1.4.  

\section{Conclusions}\label{sec:Conclusion}
A major challenge to realizing the direct wholesale market participation of DERA is enabling profit-maximizing DERAs to effectively compete with the retail programs offered by utilities in the distribution networks. To this end, this paper considers the {\em competitive DER aggregation} of a profit-seeking DERA in the wholesale electricity market. As a wholesale market participant, DERA can both inject and withdraw power from the wholesale market. It is shown that the proposed DERA model maximizes its profit while providing competitive services to its customers with higher surpluses than those offered by the distribution utilities. We also establish that the resulting social welfare from DERA's participation on behalf of its prosumers is the same as that gained by the direct participation of price-taking prosumers, making the proposed DERA aggregation model optimal in achieving wholesale market efficiency. Additionally, we derive two significant optimal price-quantity bids of DERA, of which one is submitted to the wholesale market, and the other to the distribution network access allocation \cite{ChenBoseMountTong23DERA}. 

{An open issue for future research in the proposed aggregation framework is that prosumer payment functions are nonlinear in electricity quantities and non-uniform across customers.  Although each customer is guaranteed to be better off than the competing scheme, two customers producing the same amount may be paid and compensated differently. Consequently, total charges or credits depend not only on energy quantity but also on flexibility characteristics and distribution network constraints, such as line losses and three-phase imbalance. Note that a profit-seeking DERA participating in the wholesale electricity market is not subject to the same regulations as a regulated utility.  Such non-uniform pricing may be acceptable and has also been proposed in the form of non-uniform fixed charges \cite{ borenstein2021designing, GaoAlshehriBirge:22, NEM3_JointUtility}.}



{
\bibliographystyle{IEEEtran}
\bibliography{BIB}
}

\appendix 

\subsection{Participation Model of Prosumers}\label{sec:BenchmarkA}
A prosumer in a distribution system can choose to enroll in a NEM retail program offered by her utility or a DERA providing energy services. In this context, a summary of short-run analysis over several existing models for the participation of prosumers in the regulated utility and different DERA schemes is presented. 


\subsubsection{NEM benchmarks}\label{sec:NEMsurplus}
Considering the benchmark performance of a regulated utility offering the NEM tariff, we extend the results in \cite{AlahmedTong2022NEM,alahmed2022integrating} and present closed-form characterizations of consumer/prosumer surpluses. 

For simplicity, we consider one representative prosumer by setting ${\cal N}_m=1, K=1$ and dropping the prosumer index $n$ and PoA index $m$. The prosumer's net consumption is 
\begin{equation}
   z := d-g,\nn
\end{equation}
where $g\in [0, \infty)$ is the BTM distributed generation (DG).  The prosumer is a producer if $z<0$ and a consumer if $z\ge 0$.

In evaluating the benchmark prosumer surplus under a regulated utility, we assume that the prosumer maximizes its surplus under the utility's NEM tariff, where $\pi^+$ is the retail (consumption) rate, $\pi^-$ the sell (production) rate, and $\pi^0$ the connection charge. In general $\pi^-\le \pi^+$ under NEM tariff, and the prosumer's energy bill  $P(z)$ for the net consumption $z$ is given by the following convex function
\bea \label{eq:model}
P(z) := \max\{\pi^+ z, \pi^-z\} + \pi^{\mbox{\tiny 0}}.\nn
\eea
The prosumer surplus under NEM is  $$S(d):=U(d)-P(z).$$

For an {\em active prosumer} whose consumption is a function of the available DG output $g$, the optimal consumption $d_{\mbox{\tiny NEM-a}}$ and prosumer surplus $S_{\mbox{\tiny NEM-a}}(g)$ can be obtained by
\beq\begin{array}{l}
d_{\mbox{\tiny NEM-a}} := \underset{d\in {\cal D}}{\arg\max} \bigg(U(d)-P(d-g)\bigg).\nn
\end{array}
\eeq
For the fairness of comparison,  we assume the aggregated customer is subject to the same distribution network injection and withdrawal access limits, \ie $-\underline{C} \le g-d \le \overline{C}$, which is the same as that applied to the proposed DERA optimization (\ref{eq:DERAsurplus_LnGPCC}). So, for the above optimization, the domain  is
\beq \label{eq:Ddomain}
{\cal D}:=[ \max\{\underline{d},g-\overline{C} \} , \min\{\bar{d},g+\underline{C}\}].\nn
\eeq

The surplus $S_{\mbox{\tiny NEM-a}}$  and the consumption $d_{\mbox{\tiny NEM-a}}$ of an active prosumer are given by the following equations.
\begin{align}\label{eq:acitiveSurplus}
    &S_{\mbox{\tiny NEM-a}}(g, \underline{C}, \overline{C})=U(d_{\mbox{\tiny NEM-a}})-P(d_{\mbox{\tiny NEM-a}}-g)\\ &=\begin{cases}
U(d^-)-\pi^-(d^--g)-\pi^0,& g\geq d^-\nn\\
U(d^+)-\pi^+(d^+-g)-\pi^0,&  g \le  d^+\nn\\
U(d^0)-\pi^0, &\text{otherwise}\\
\end{cases}\\
&d_{\mbox{\tiny NEM-a}}= \max \{d^+,\min \{g,d^-\}\},\nn
\end{align}
where we denote
\beq \label{eq:fconsump}
\begin{array}{c}
d^+:=f(\pi^+),  ~~d^-:=f(\pi^-), ~~d^0:=f(\mu^*(g))\\
f(x):=\max \{\underline{d}, g-\overline{C}, \min \{V^{-1}(x),\bar{d}, g+\underline{C}\}\}, 
\end{array}
\eeq
and, by solving $f(\mu)=g$, we have $\mu^*(g)\in [\pi^-,\pi^+]$. 

A prosumer is called passive if it decides energy consumption without the awareness of its DG output and the influence brought by NEM switching among $\pi^-$ and $\pi^+$. The optimal consumption bundle of such a  {\em passive prosumer} under the NEM tariff is given by 
\beq\begin{array}{l}\label{eq:passiveD}
d_{\mbox{\tiny NEM-p}} := \underset{d\in {\cal D}}{\arg\max}  \bigg(U(d)-\pi^+d\bigg).\nn
\end{array}
\eeq
The total consumption $d_{\mbox{\tiny NEM-p}}  $ and the surplus $S_{\mbox{\tiny NEM-p}}$ of a  {\em passive prosumer} are given by
\begin{align}\label{eq:passiveSurplus}
    &S_{\mbox{\tiny NEM-p}}(g, \underline{C}, \overline{C})=U(d_{\mbox{\tiny NEM-p}})-P(d_{\mbox{\tiny NEM-p}}-g)\\ &=\begin{cases}
U(d^+)-\pi^-(d^+-g)-\pi^0,&g \geq d^+\nn\\
U(d^+)-\pi^+(d^+-g)-\pi^0,& g<d^+ 
\end{cases}\\
&d_{\mbox{\tiny NEM-p}}=  d^+.\label{eq:dNEMp}
\end{align}
In practice, because active prosumer decision requires installing special DG measurement devices and sophisticated control, most prosumers are passive.\footnote{Britain establishes a database for passive customers and encourages the participation of passive customers in the electricity market \cite{Ros18report}.} In summary, the prosumer surplus under NEM, $S_{\mbox{\tiny NEM}}(g, \underline{C}, \overline{C})$ is given by
\begin{align}\label{eq:NEMS}
S_{\mbox{\tiny NEM}}(g, \underline{C}, \overline{C})=\begin{cases}
S_{\mbox{\tiny NEM-a}}(g, \underline{C}, \overline{C}),& \text{active prosumer},\\
S_{\mbox{\tiny NEM-p}}(g, \underline{C}, \overline{C}),& \text{passive prosumer}. 
\end{cases}
\end{align}

\subsection{Proof of Theorem~\ref{thm:DERA} and Proposition \ref{Prop:DERA}}\label{sec:DERADC}

We prove this proposition using Karush-Kuhn-Tucker (KKT) conditions of convex optimization (\ref{eq:DERAsurplus_LnGPCC}), based on feasibility and utility function assumptions in Sec. \ref{sec:closeform}. 

Assign dual variables to (\ref{eq:DERAsurplus_LnGPCC}), we have 
\bea\label{eq:DERAsurplus_compete} 
\begin{array}{lrl}
&\underset{\omegabf, \mathsf{D}}{\rm max}&~~ \sum_{n=1}^N (\omega_n-\pi (\mathbf{1}^\top \dbf_n-g_n))\\
&  \mbox{subject to} &  \forall n \in [N], \forall m \in [M], \\
&(\underline{\gamma}_m,\overline{\gamma}_m):& -\underline{C}_m \le \sum_{n=1}^{{\cal N}_m}  (g_n-\mathbf{1}^\top \dbf_n) \le \overline{C}_m,\\
&\chi_{n}:&  U_n(\dbf_n)-\omega_n \geq {\cal K}_n, \\
& (\underline{\nubf}_{n},\overline{\nubf}_{n}):&\underline{\dbf}_n \preceq  \dbf_n \preceq  \bar{\dbf}_n,
\end{array}
\eea

The Lagrangian function is 
\beq
\begin{array}{lrl}
{\cal L}(\cdot)=- \sum_{n=1}^N (\omega_n-\pi(\mathbf{1}^\top \dbf_n-g_n))\\
~~~~~~+\sum_{n=1}^N\chi_n({\cal K}_n - \sum_{k=1}^K U_{nk}(d_{nk})+\omega_n)\\
~~~~~~+\sum_{n=1}^N\underline{\nubf}_n^\top(\underline{\dbf}_n - \dbf_n)+ \sum_{n=1}^N\overline{\nubf}_n^\top( \dbf_n- \overline{\dbf}_n)\\
~~~~~~+\sum_{m=1}^M\underline{\gamma}_m(-\underline{C}_m- \sum_{n=1}^{{\cal N}_m}  (g_n-\mathbf{1}^\top \dbf_n))\\
~~~~~~+\sum_{m=1}^M\overline{\gamma}_m (\sum_{n=1}^{{\cal N}_m}  (g_n-\mathbf{1}^\top \dbf_n)- \overline{C}_m).
\end{array}
\eeq
Hence, from KKT conditions of (\ref{eq:DERAsurplus_compete}),  $\forall n \in [N], \forall m \in [M]$,
\bea\label{eq:DERA_KKT}
\begin{array}{lrl}
\frac{\partial \cal L}{\partial \omega_n}&=&\chi^*_{n}-1=0 \Rightarrow \chi^*_{n}=1\overset{(a)}{\Rightarrow} \omega^{*}_n = U_n(\dbf_n^*)-{\cal K}_n \\
\frac{\partial \cal L}{\partial d_{nk}}&=&\pi -\chi_{n}^*V_{nk}(d_{nk}^*)-\underline{\nu}_{nk}^*+\overline{\nu}_{nk}^* +\underline{\gamma}^*_m-\overline{\gamma}^*_m \\&=&\pi -V_{nk}(d_{nk}^*)-\underline{\nu}_{nk}^*+\overline{\nu}_{nk}^* +\underline{\gamma}^*_m-\overline{\gamma}^*_m =0.
\end{array}
\eea
where superscript * represents the optimal solution. (a) comes from the complementary slackness condition 
\beq
\chi^*_{n}(\omega^{*}_n - U_n(\dbf_n^*)+{\cal K}_n)=\omega^{*}_n - U_n(\dbf_n^*)+{\cal K}_n=0.
\eeq
So, \eqref{eq:omega*} is proved.

We complete the proof by considering three cases:  (1) $\sum\limits_{n=1}^{{\cal N}_m} g_n \geq \sum\limits_{n=1}^{{\cal N}_m}\sum\limits_{k=1}^K h_{nk}(\pi)+\overline{C}_m$, (2) $\sum\limits_{n=1}^{{\cal N}_m} g_n  \le \sum\limits_{n=1}^{{\cal N}_m}\sum\limits_{k=1}^K h_{nk}(\pi) -\underline{C}_m$, and (3)  all remaining cases.

\underline{(1) When $\sum\limits_{n=1}^{{\cal N}_m} g_n \geq \sum\limits_{n=1}^{{\cal N}_m}\sum\limits_{k=1}^K h_{nk}(\pi)+\overline{C}_m$}, we have
  
\beq 
\begin{array}{lrl}
\sum\limits_{n=1}^{{\cal N}_m}\sum\limits_{k=1}^K h_{nk}(\overline{\xi}_m)&:=&\sum\limits_{n=1}^{{\cal N}_m}  g_n-\overline{C}_m\geq \sum\limits_{n=1}^{{\cal N}_m}\sum\limits_{k=1}^Kh_{nk}(\pi),\\
&\overset{(b)}{\Rightarrow} &\pi \geq \overline{\xi}.\label{eq:UBxiPi}
\end{array}
\eeq
(b) comes from the concavity of the utility function, indicating $V_{nk}^{-1}(\cdot)$ thus $h_{nk}(\cdot)$ in \eqref{eq:h} nonincreasing.

Let $\underline{\gamma}^*_m=0, \overline{\gamma}^*_m=\pi-\overline{\xi}_m$, $d^*_{nk}= 
    h_{nk}(\overline{\xi}_m)$, and 
    \beq  
\begin{array}{l}
(\underline{\nu}_{nk}^*, \overline{\nu}_{nk}^*)=\begin{cases}
    (\overline{\xi}_m-V_{nk}(\underline{d}_{nk}),0), &  \underline{d}_{nk} \geq V_{nk}^{-1}(\overline{\xi}_m) \\
    (0,V_{nk}(\overline{d}_{nk})-\overline{\xi}_m), &   V_{nk}^{-1}(\overline{\xi}_m)\geq \overline{d}_{nk} \\
     (0,0), & \text{otherwise}
    \end{cases}.\nn
\end{array}
\eeq
We find all KKT conditions including \eqref{eq:DERA_KKT}, the primal and dual constraints are satisfied.

 The optimal value can be computed by
\beq
\begin{array}{l}
\Pi(\bar{\Cbf},\underline{\Cbf})=\sum\limits_{n=1}^N (\omega_{n}^*-\pi (\sum_{k=1}^K d_{nk}^*-g_n))\\
    ~~=\sum\limits_{m=1}^M[\sum\limits_{n=1}^{{\cal N}_m}\big(U_n( \hbf_n(\overline{\xi}_m))-{\cal K}_n-\pi(\hbf_n(\overline{\xi}_m)-g_n)\big)]\\
~~=\sum\limits_{m=1}^M[\sum\limits_{n=1}^{{\cal N}_m}U_n( \hbf_n(\overline{\xi}_m))-\pi\sum\limits_{n=1}^{{\cal N}_m}(\hbf_n(\overline{\xi}_m)-g_n)]-\sum\limits_{n=1}^N{\cal K}_n\\
~~=\sum\limits_{m=1}^M[\sum\limits_{n=1}^{{\cal N}_m}U_n( \hbf_n(\overline{\xi}_m))+\pi\overline{C}_m]-\sum\limits_{n=1}^N{\cal K}_n,\nn
\end{array}
\eeq
where $\hbf_n(x) := \sum_{k=1}^K h_{nk}(x)$, and $\sum\limits_{m=1}^M\sum\limits_{n=1}^{{\cal N}_m} 1 = \sum\limits_{n=1}^N 1$ from the definition of ${\cal  N}_m$ in\eqref{eq:N}. So the optimal solution and optimal value satisfy formulations in \eqref{eq:d*} of Theorem~\ref{thm:DERA} and \eqref{eq:bidCurveall} of Proposition \ref{Prop:DERA}.

\underline{(2) When $\sum\limits_{n=1}^{{\cal N}_m} g_n  \le \sum\limits_{n=1}^{{\cal N}_m}\sum\limits_{k=1}^K h_{nk}(\pi) -\underline{C}_m$}, we  have 
\beq \label{eq:LBxiPi}
\underline{\xi}_m\geq \pi. 
\eeq
Let $\overline{\gamma}^*_m =0, \underline{\gamma}^*_m=\underline{\xi}_m -\pi$, $d^*_{nk}= 
    h_{nk}(\underline{\xi}_m)$, and 
    \beq  
\begin{array}{l}
(\underline{\nu}_{nk}^*, \overline{\nu}_{nk}^*)=\begin{cases}
    (\underline{\xi}_m-V_{nk}(\underline{d}_{nk}),0), &  \underline{d}_{nk} \geq V_{nk}^{-1}(\underline{\xi}_m) \\
    (0,V_{nk}(\overline{d}_{nk})-\underline{\xi}_m), &   V_{nk}^{-1}(\underline{\xi}_m)\geq \overline{d}_{nk} \\
     (0,0), & \text{otherwise}
    \end{cases}.\nn
\end{array}
\eeq
We find all KKT conditions including \eqref{eq:DERA_KKT}, the primal and dual constraints are satisfied.

 The optimal value can be computed by
\beq
\begin{array}{l}
\Pi(\bar{\Cbf},\underline{\Cbf})=\sum\limits_{n=1}^N (\omega_{n}^*-\pi (\sum_{k=1}^K d_{nk}^*-g_n))\\
    ~~~~=\sum\limits_{m=1}^M[\sum\limits_{n=1}^{{\cal N}_m}\big(U_n( \hbf_n(\underline{\xi}_m))-{\cal K}_n-\pi(\hbf_n(\underline{\xi}_m)-g_n)\big)]\\
~~~~=\sum\limits_{m=1}^M[\sum\limits_{n=1}^{{\cal N}_m}U_n( \hbf_n(\underline{\xi}_m))-\pi\underline{C}_m]-\sum\limits_{n=1}^N{\cal K}_n.\nn
\end{array}
\eeq 
So the optimal solution and optimal value satisfy formulations in \eqref{eq:d*} of Theorem~\ref{thm:DERA} and \eqref{eq:bidCurveall} of Proposition \ref{Prop:DERA}.

\underline{(3) When $\sum\limits_{n=1}^{{\cal N}_m}\hbf_n(\pi) -\underline{C}_m <  \sum\limits_{n=1}^{{\cal N}_m} g_n <  \sum\limits_{n=1}^{{\cal N}_m}\hbf_n(\pi)+\overline{C}_m $}, let $\overline{\gamma}^*_m =0, \underline{\gamma}^*_m=0$, $d^*_{nk}= 
    h_{nk}(\pi)$, and 
    \beq 
\begin{array}{l}
(\underline{\nu}_{nk}^*, \overline{\nu}_{nk}^*)=\begin{cases}
    (\pi-V_{nk}(\underline{d}_{nk} ),0), &  \underline{d}_{nk} \geq V_{nk}^{-1}(\pi) \\
    (0,V_{nk}(\overline{d}_{nk})-\pi), &   V_{nk}^{-1}(\pi)\geq \overline{d}_{nk} \\
     (0,0), & \text{otherwise}
    \end{cases}.\nn
\end{array}
\eeq
We find all KKT conditions including \eqref{eq:DERA_KKT}, the primal and dual constraints are satisfied. 
The optimal value is given by the equation below, which is not a function of $(\underline{\Cbf},\overline{\Cbf})$.
\begin{align}
    \begin{aligned}
    &\Pi=\sum_{n=1}^N(U_{n}(\hbf_n(\pi))-\pi(\hbf_n(\pi)-g_n))-\sum\limits_{n=1}^N{\cal K}_n.
    \end{aligned}
\end{align}
So the optimal solution and optimal value satisfy formulations in \eqref{eq:d*} of Theorem~\ref{thm:DERA} and \eqref{eq:bidCurveall} of Proposition \ref{Prop:DERA}. \hfill\QED

\subsubsection{Two-part pricing in GAB}

We summarize below the optimal DERA two-part pricing scheme in \cite{GaoAlshehriBirge:22}, aggregating BTM DG productions. The original pricing scheme keeps the customer surplus under DERA competitive with that when the customers directly buy energy from the wholesale market. Here, considering the realistic retail market setting, we revised the DERA pricing model to be competitive with that when the customers directly buy energy from the incumbent utility company under NEM.

The two-part pricing includes a variable price $\lambda_i$ and a discriminative fixed charge $\delta_i$. Prosumers can sell energy $x_i$ to the DERA with price $\lambda_i$, and buy energy from the energy provider, e.g. the utility company, with the retail rate $\pi^+$.  In this case, the surplus maximization of  prosumer $i$ is given by
\beq
\begin{array}{c}
\underset{d_i\in {\cal D}_i, x_i \in [0,g_i]}{\max}~  S_i^{\mbox{\tiny Pro}
}(\cdot) ,
\end{array}
\eeq
 where ${\cal D}_i:=[ \max\{\underline{d}_i,g_i-\overline{C}_i \} , \min\{\bar{d}_i,g_i+\underline{C}_i\}]$, and
$$
 S_i^{\mbox{\tiny Pro}
}(\cdot)=\begin{cases} U_i(d_i)- \pi^+ [d_i-g_i+x_i]^++\lambda_i x_i -\delta_i,& x_i >0\\ 
U_i(d_i)-\pi^+[d_i-g_i]^+,&x_i= 0\\ \end{cases}.
$$
 With function $f_i$ defined in (\ref{eq:fconsump}), the optimal energy consumption of  DERA  computed from the optimization above is given by 
$$
 \begin{cases}
     x^*_i(\lambda_i, \delta_i)=[g_i-f_i(\lambda_i)]^+,~ d^*_i(\lambda_i, \delta_i)=g_i-x^*_i,\\
     x^*_i(\lambda_i, \delta_i)=0, ~d^*_i(\lambda_i, \delta_i)= g_i+[d^+_i-g_i]^+,
 \end{cases}
$$
 where  $d^+_i:=f_i(\pi^+)$.
If prosumer chooses not to sell energy to DERA, the maximum prosumer surplus is given by
$$
 S^{\mbox {\tiny NO}}_i=\begin{cases}
     U_i(f_i(0)),&\text{if }~ f_i(0) \le g_i, \\
     U_i(g_i), &\text{if }~  d_i^+ \le g_i < f_i(0),\\
     U_i(d_i^+)-\pi^+ (d_i^+-g_i), &\text{if }~ 0 \le g_i < d_i^+.
     \end{cases}
$$

 To make DERA competitive with the incumbent utility company, we set ${\cal K}_i=\zeta S^{\mbox {\tiny NO}}_i$ for the ${\cal K}$-competitive constraint. That way, the prosumer selling $x^*_i$ energy to DERA will always have $\zeta$ times its surplus under the incumbent utility company with NEM.
 
 The profit maximization of the DERA when participating in the wholesale market with price $\pi$ is given by 
\begin{equation}\label{twoDERA} \begin{array}{lrl}&\underset{\{\lambda_i, \delta_i\} }{\rm max} & \sum_i (\delta_i \mathbbm{1}\{g_i-f_i(\lambda_i)>0\}+\\
&&~~~~~~~~~~~~~~~~(\pi -\lambda_i)[g_i-f_i(\lambda_i)]^+ )\\
& s.t. &
{\cal K}_i\le  U_i(f_i(\lambda_i))+\lambda_i[g_i-f_i(\lambda_i)]^+-\delta_i,\\
&& \lambda_i \le \pi.
\end{array}\end{equation}
The optimal pricing from the above optimization is 
$$
\begin{cases}
\lambda_i^* = \pi , \forall i,\\
\delta_i^*=U_i(f_i(\lambda_i^*))+\lambda_i^*[g_i-f_i(\lambda^*_i)]^+-{\cal K}_i.
\end{cases}
$$
So, when $g_i-f_i(\pi)> 0$, the prosumer will be aggregated by DERA for its  extra BTM DG generation. Otherwise, the prosumer will stay under the  utility company with NEM. And the customer surplus under this two-part pricing  has 
\beq 
S_i^{\mbox{\tiny GAB}}={\cal K}_i=\zeta S^{\mbox {\tiny NO}}_i.
\eeq
In \cite{GaoAlshehriBirge:22}, GAB also needs to set $\zeta$ carefully to avoid the DERA deficit as in Proposition~\ref{prop:RA}.

 \vspace{-0.3cm}
\subsection{Proof of Proposition~\ref{prop:Pcappi+}}\label{sec:prop1} 
We prove Proposition~\ref{prop:Pcappi+} by considering four cases:  (1) $g_n \le d_n^+$ for both active and passive prosumers, (2) $g_n > d_n^+$ for a passive prosumer, (3) $ d_n^+< g_n \le d_n^-$ for an active prosumer, and (4) $ g_n > d_n^-$ for an active prosumer.

 \underline{(1) If $g_n \le d_n^+$},   we have
\begin{align}
    \begin{aligned}
\omega^{*}_n &\overset{(a)}{=}U_n(d_n^*)-{\cal K}_n\\
&\overset{(b)}{=}U_n(d_n^*)-\zeta S_n^{\mbox{\tiny NEM}}(g_n, \underline{C}_n, \overline{C}_n)\\
&\overset{(c)}{\le} U_n(d_n^*)-S_n^{\mbox{\tiny NEM}}(g_n, \underline{C}_n, \overline{C}_n)\nn\\
&\overset{(d)}{\le} U_n(d_n^*)-(U_n(d_n^+)-\pi^+d_n^+)\nn\\
&\overset{(e)}{\le}  d_n^*\pi^+. 
     \end{aligned}
\end{align}
Here, (a) comes from the optimal solution in Theorem~\ref{thm:DERA}. (b) relies on the setting that ${\cal K}_n =\zeta S_n^{\mbox{\tiny NEM}}(g_n, \underline{C}_n, \overline{C}_n)$ in \eqref{eq:NEMK}. (c) replies on
$\zeta\geq 1$ and the nonnegative prosumer surplus assumption $S_n^{\mbox{\tiny NEM}}\geq 0$ at the beginning of Sec.~\ref{sec:CNEM}. (d) comes from the definition of $S_n^{\mbox{\tiny NEM}}$ given in \eqref{eq:NEMS}.\footnote{We ignore the fix charge $\pi^0$ under NEM X here for simplicity.}(e) comes from 
\beq\label{eq:d+OPT}
U_n(d_n^*)- d_n^*\pi^+ \le U_n(d_n^+)-\pi^+d_n^+,\eeq
which can be derived from the optimality of $$d^+:={\rm arg~max}_{ d \in {\cal D}} (U(d)-\pi^+d)$$ 
with domain defined in \eqref{eq:Ddomain}.

\underline{(2) If $g_n > d_n^+$, for a passive prosumer},  we have
\begin{align}
    \begin{aligned}
\omega^{*}_n &\overset{(a)}{=}U_n(d_n^*)-{\cal K}_n\\
&\overset{(b)}{=}U_n(d_n^*)-\zeta S_n^{\mbox{\tiny NEM}}(g_n, \underline{C}_n, \overline{C}_n)\\
&\overset{(c)}{\le} U_n(d_n^*)-S_n^{\mbox{\tiny NEM}}(g_n, \underline{C}_n, \overline{C}_n)\nn\\
&\overset{(d)}{\le} U_n(d_n^*)- (U_n(d_n^+)-\pi^-(d_n^+-g_n)) \nn\\
&\overset{(e)}{\le} U_n(d_n^*)- U_n(d_n^+) \nn\\
&\overset{(f)}{\le} \pi^+ (d_n^*-d_n^+)\nn\\ 
&\overset{(g)}{\le}  d_n^*\pi^+.  
     \end{aligned}
\end{align}
Here,(a) comes from the optimal solution in Theorem~\ref{thm:DERA}. (b) relies on the setting that ${\cal K}_n =\zeta S_n^{\mbox{\tiny NEM}}(g_n, \underline{C}_n, \overline{C}_n)$ in \eqref{eq:NEMK}. (c) replies on
$\zeta\geq 1$ and the assumption that $S_n^{\mbox{\tiny NEM}}\geq 0$. (d) comes from the definition of $S_n^{\mbox{\tiny NEM}}$ given in \eqref{eq:NEMS}.  (e) replies on $g_n > d_n^+$ and $\pi^-\geq 0$. (f) comes from  \eqref{eq:d+OPT}. (g) holds because $d_n^+\geq 0$ and $\pi^+\geq 0$.

\underline{(3) If $ d_n^+< g_n \le d_n^-$, for an active prosumer},  we have
\begin{align}
    \begin{aligned}
\omega^{*}_n   
&\overset{(a)}{\le} U_n(d_n^*)-S_n^{\mbox{\tiny NEM}}(g_n, \underline{C}_n, \overline{C}_n)\nn\\
&\overset{(b)}{\le} U_n(d_n^*)- U_n(d_n^0) \nn\\
&\overset{(c)}{\le} 0 \le d_n^*\pi^+.  
     \end{aligned}
\end{align}
Here, (a) comes from Theorem~\ref{thm:DERA} and the assumption  $S_n^{\mbox{\tiny NEM}}\geq 0$. (b) comes from the definition of $S_n^{\mbox{\tiny NEM}}$ given in \eqref{eq:NEMS} when $ d_n^+< g_n \le d_n^-$ for active prosumer. (c) comes from the optimality of $d^0:={\rm arg~max}_{ d \in {\cal D}} U(d)$, $d_n^*\geq 0$, and $\pi^+\geq  0$.

\underline{(4) If $ g_n > d_n^-$, for an active prosumer},  we have
\begin{align}
    \begin{aligned}
\omega^{*}_n   
&\overset{(a)}{\le} U_n(d_n^*)-S_n^{\mbox{\tiny NEM}}(g_n, \underline{C}_n, \overline{C}_n)\nn\\
&\overset{(b)}{\le} U_n(d_n^*)- (U_n(d_n^-)-\pi^-(d_n^--g_n)) \nn\\
&\overset{(c)}{\le} U_n(d_n^*)- U_n(d_n^-) \nn\\
&\overset{(d)}{\le} \pi^- (d_n^*-d_n^-)\nn\\ 
&\overset{(e)}{\le}  d_n^*\pi^- \le d_n^*\pi^+. 
     \end{aligned}
\end{align}
Here, (a) comes from Theorem~\ref{thm:DERA} and the assumption  $S_n^{\mbox{\tiny NEM}}\geq 0$. (b) comes from the definition of $S_n^{\mbox{\tiny NEM}}$ given in \eqref{eq:NEMS} when $g_n \geq d^-_n$.  (c) holds because $g_n \geq d^-_n$. (d) comes from    
\beq\label{eq:d-OPT}
U_n(d_n^*)- d_n^*\pi^- \le U_n(d_n^-)-\pi^-d_n^-,\eeq
which can be derived from the optimality of $$d^-:={\rm arg~max}_{ d \in {\cal D}} (U(d)-\pi^-d).$$ 
(e) holds because $d_n^-\geq 0$ and $\pi^+\geq \pi^-\geq 0$. \hfill\QED

\subsection{Proof of Proposition~\ref{prop:RA}} \label{sec:prop2}

We compute the expected DERA surplus by 

\begin{align}
    \begin{aligned}
\mathbb{E}_{\pi}&[\Pi_{\mbox{\tiny DERA}}| \gbf]\overset{(a)}{=}  \mathbb{E}_{\pi}[\sum_{n=1}^N (\omega^*_n-\pi(d_n^*-g_n))]\\
& \overset{(b)}{=}  \sum_{n=1}^N \mathbb{E}_{\pi} [U_n(d^{*}_n)-{\cal K}_n-\pi(d_n^*-g_n)]\\
& \overset{(c)}{=} \sum_{n=1}^N (\mathbb{E}_{\pi}[U_n(d^{*}_n)-\pi(d_n^*-g_n)]-\zeta S_n^{\mbox{\tiny NEM}}(g_n, \underline{C}_n, \overline{C}_n)),\\ 
&  \mathbb{E}_{\pi, \gbf}[\Pi_{\mbox{\tiny DERA}}] \overset{(d)}{=} \mathbb{E}_{\gbf}\bigg[\mathbb{E}_{\pi}[\Pi_{\mbox{\tiny DERA}}|\gbf]\bigg] \\
&\overset{(e)}{=} \sum_{n=1}^N (\mathbb{E} [U_n(d^{*}_n)-\pi(d_n^*-g_n)]-\zeta \mathbb{E} [S_n^{\mbox{\tiny NEM}}(g_n, \underline{C}_n, \overline{C}_n)])\\&\overset{(f)}{\geq} 0. \\ \nn
     \end{aligned}
\end{align}
Here, (a) comes from the definition of DERA surplus, which is the objective function of (\ref{eq:DERAsurplus_LnGPCC}) and (b) comes from the optimal solution in Theorem~\ref{thm:DERA}. (c) comes from \eqref{eq:NEMK} when DERA is competitive with NEM. (d) relies on the independence of BTM DG generation $\gbf$ and wholesale LMP $\pi$ in a competitive market. (e) brings in the formulation of $\mathbb{E}_{\pi}[\Pi_{\mbox{\tiny DERA}}| \gbf]$. (f) comes from summing equation \eqref{eq:Udzeta} below for $\forall n\in [N]$. 
\beq\label{eq:Udzeta}
\mathbb{E}_{\pi, \gbf}[U_n(d_n^*)-\pi  (d^*_{n}-g_n)] \geq   \zeta \mathbb{E}_{\gbf}[S_n^{\mbox{\tiny NEM}}(g_n, \underline{C}_n, \overline{C}_n)].
\eeq
The proof of \eqref{eq:Udzeta} follows the assumption $\mathbb{E}_{g_n}[S_n^{\mbox{\tiny NEM}}] \geq 0, \forall n \in [N]$ at the beginning of Sec.~\ref{sec:CNEM}, which represents nonegative  prosumers' surpluses on expectation under NEM. 

When $\mathbb{E}_{g_n}[S_n^{\mbox{\tiny NEM}}]=0$, $\mathbb{E}_{\pi}[U_n(d_n^*)-\pi  (d^*_{n}-g_n)|g_n] \geq 0$ from Lemma~\ref{lemma:DERAsurplus}. So,  based on the independency of $\pi$ and $\gbf$, \eqref{eq:Udzeta} holds. 

When $\mathbb{E}_{g_n}[S_n^{\mbox{\tiny NEM}}] > 0$, we have $\forall n\in [N]$,
\begin{align}
    \begin{aligned}
    \frac{\mathbb{E}_{\pi, \gbf}[U_n(d_n^{*})-\pi(d^{*}_{n}-g_n)]}{\mathbb{E}_{ \gbf}[S_n^{\mbox{\tiny NEM}}(g_n, \underline{C}_n, \overline{C}_n)]} &\geq \underset{n}{min} \frac{\mathbb{E}_{\pi, \gbf}[U_n(d_n^{*})-\pi (d^{*}_{n}-g_n)]}{\mathbb{E}_{\gbf}[S_n^{\mbox{\tiny NEM}}(g_n, \underline{C}_n, \overline{C}_n)]} \\
    &\overset{(a)}{\geq} min_{n \in [N]} ~\overline{\zeta}_n \overset{(b)}{\geq} \zeta \Rightarrow \eqref{eq:Udzeta}.
         \end{aligned}
\end{align}
Here (a) comes from the upper bound defined by  $$\overline{\zeta}_n:=\mathbbm{1}_{\{S_n^{\mbox{\tiny NEM}} >0\}} \mathbb{E}[U_n(d_n^{*})-\pi  (d^{*}_{n}-g_n)]/\mathbb{E}[S_n^{\mbox{\tiny NEM}}]+\mathbbm{2}_{\{S_n^{\mbox{\tiny NEM}} =0\}}.$$ 
From Lemma~\ref{lemma:DERAsurplus} and the independency of $\pi$ and $\gbf$, we know $\overline{\zeta}_n \geq 1, \forall n \in [N]$. Therefore, (b) follows the existence of $\zeta \in [1, min_{n \in [N]} ~\overline{\zeta}_n]$. 

\hfill\QED

\begin{lemma}[] \label{lemma:DERAsurplus}
Suppose NEM and LMP has $\pi^-\le \mathbb{E}[\pi]\le \pi^+$, then given $\gbf:=(g_n)$, we have
\beq 
\mathbb{E}_{\pi}[U_n(d_n^*)-\pi  (d^*_{n}-g_n)] \geq   S_n^{\mbox{\tiny NEM}}(g_n, \underline{C}_n, \overline{C}_n), \forall n \in [N].
\eeq 
\end{lemma}

Proof: $d^+_n$, $d^-$, and $d^0$ are  defined with \eqref{eq:fconsump}.

\underline{When $d^+_n \geq g_n$}, we have  
\begin{align}
    \begin{aligned}
    \mathbb{E}_{\pi}[U_n(d_n^*)-\pi  (d^*_{n}-g_n)]&\overset{(a)}{\geq} \mathbb{E}_{\pi}[U_n(d_n^+)-\pi(d_{n}^+-g_n)]\\
    &\overset{(b)}{\geq} U_n(d_n^+)-\mathbb{E}[\pi](d_{n}^+-g_n)\\
    &\overset{(c)}{\geq}   U_n(d_n^+)-\pi^+ (d_{n}^+-g_n) \\
    &\overset{(d)}{\geq} S_n^{\mbox{\tiny NEM}}(g_n, \underline{C}_n, \overline{C}_n).\nn
     \end{aligned}
\end{align}
Here, (a) follows the optimality of 
\beq\begin{array}{l}\label{eq:passiveLMP}
d^* _n(\pi):= \arg\max_{d_n \in {\cal D}_n} \bigg(U_n(d_n)-\pi(d_n -g_n)\bigg)
\end{array}
\eeq
for all realizations of $\pi$. Here, 
\beq
{\cal D}_n:=[ \max\{\underline{d}_n,g_n-\overline{C}_n \} , \min\{\bar{d}_n,g_n+\underline{C}_n\}].
\eeq
(b) follows the linearity of expectation. (c) relies on  the condition $ \mathbb{E}[\pi]\le \pi^+$ and $d^+_n \geq g_n$. (d) comes from  the definition of $S_n^{\mbox{\tiny NEM}}(g_n, \underline{C}_n, \overline{C}_n)$ in \eqref{eq:NEMS} and $\pi^0\geq 0$.

\underline{When this prosumer is passive and $d^+_n < g_n$}, we have  
\begin{align}
    \begin{aligned}
    \mathbb{E}_{\pi}[U_n(d_n^*)-\pi    (d^*_{n}-g_n)]&\overset{(a)}{\geq} U_n(d_n^+)-\mathbb{E}[\pi] (d_{n}^+-g_n)\\
    &\overset{(b)}{\geq}   U_n(d_n^+)-\pi^- (d_{n}^+-g_n) \\
    &\overset{(c)}{\geq} S_n^{\mbox{\tiny NEM}}(g_n, \underline{C}_n, \overline{C}_n),\nn
     \end{aligned}
\end{align}
where (a) follows the optimality of \eqref{eq:passiveLMP} for all $\pi$, (b) comes from the condition that $\pi^-\le \mathbb{E}[\pi]$ and $d^+_n < g_n$, and (c) comes from  the definition of $S_n^{\mbox{\tiny NEM}}(g_n, \underline{C}_n, \overline{C}_n)$ in \eqref{eq:NEMS}.

\underline{When this prosumer is active and $d^+_n < g_n < d^-_n$}, we have
\begin{align}
    \begin{aligned}
    \mathbb{E}_{\pi}[U_n(d_n^*)-\pi (d^*_{n}-g_n)]&\overset{(a)}{\geq} U_n(d_n^0)-\mathbb{E}[\pi] (d_{n}^0-g_n)\\
    &\overset{(b)}{\geq}   U_n(d_n^0) \\
    &\overset{(c)}{\geq} S_n^{\mbox{\tiny NEM}}(g_n, \underline{C}_n, \overline{C}_n),\nn
     \end{aligned}
\end{align}
where (a) follows the optimality of \eqref{eq:passiveLMP} for all $\pi$, (b) follows $d_{n}^0=g_n$ with \eqref{eq:fconsump}, and (c) comes from the definition in \eqref{eq:NEMS}.

\underline{When this prosumer is active and $g_n \geq d^-_n$}, we have
\begin{align}
    \begin{aligned}
   \mathbb{E}_{\pi}[ U_n(d_n^*)-\pi (d^*_{n}-g_n)]&\overset{(a)}{\geq} U_n(d_n^-)-\mathbb{E}[\pi ](d^-_{n}-g_n)\\
    &\overset{(b)}{\geq}   U_n(d_n^-)-\pi^-  (d^-_{n}-g_n) \\
    &\overset{(c)}{\geq} S_n^{\mbox{\tiny NEM}}(g_n, \underline{C}_n, \overline{C}_n),\nn
     \end{aligned}
\end{align}
where (a) follows the optimality of \eqref{eq:passiveLMP} for all $\pi$, (b) relies on the condition that $\pi^-\le \mathbb{E}[\pi]$ and $d^-_n \le g_n$, and (c) comes from the definition in \eqref{eq:NEMS}. \hfill\QED

 \vspace{-0.3cm}
\subsection{Proof of Lemma~\ref{lemma:WMC}} \label{sec:lemma1} 

We here show that the LMP and dispatch result from \eqref{eq: SS_CC} is at the bidding curve of DERA, \ie \eqref{Q-P}. 

The access limits $\overline{c}_{in}$ and $\underline{c}_{in}$ represent the distribution network injection and withdrawal capacities allocated to each prosumer. We set these values the same as those from \eqref{eq:d*} to ensure a fair comparison. That way, prosumers participating directly in the wholesale market face the same distribution network access limits as those in the proposed DERA model. Specifically, solve $\overline{\xi}_m, \underline{\xi}_m$ with $K=1$ from \eqref{eq:xiLB}\eqref{eq:xiUB} and set
\beq \label{eq:AccessPropsumerUB}
\begin{array}{l}
\overline{c}_{in}:=\begin{cases}
    0, &  \sum\limits_{n=1}^{{\cal N}_m} g_{in}  \le \sum\limits_{n=1}^{{\cal N}_m} h_{in}(\pi_i) -\underline{C}_m \\
   g_{in}-h_{in}(\overline{\xi}_m ), &  \sum\limits_{n=1}^{{\cal N}_m} g_{in} \geq \sum\limits_{n=1}^{{\cal N}_m}h_{in}(\pi_i)+\overline{C}_m \\
     [g_{in}-h_{in}(\pi_i)]_+, & \text{otherwise}
    \end{cases},
\end{array}
\eeq
\beq \label{eq:AccessPropsumerLB}
\begin{array}{l}
\underline{c}_{in}:=\begin{cases}
    -g_{in}+h_{in}(\underline{\xi}_m), &  \sum\limits_{n=1}^{{\cal N}_m} g_{in}  \le \sum\limits_{n=1}^{{\cal N}_m} h_{in}(\pi_i) -\underline{C}_m \\
   0, &  \sum\limits_{n=1}^{{\cal N}_m} g_{in} \geq \sum\limits_{n=1}^{{\cal N}_m}h_{in}(\pi_i)+\overline{C}_m \\
   [-g_{in}+h_{in}(\pi_i)]_+, & \text{otherwise}
    \end{cases}.
\end{array}
\eeq

The market clearing \eqref{eq: SS_CC} is equivalent\footnote{This equivalence is proved by the convexity and matching solutions from the KKT/dual problem. The market clearing \eqref{eq: SS_CC} can be dual decomposed into \eqref{eq:DERAsurplus_LnGPCC} with ${\cal K}_{in}=0, \forall i\in [I], \forall n\in [N]$.} to the reformulation  
\beq
\begin{array}{lrl}\label{eq: SS_CCDiPro}
&\underset{\Dbf, \pbf, \ebf \geq 0}{\rm max}&~~ \sum_{i=1}^I \sum_{n=1}^N (U_{in}(d_{in})+B_i(e_i)- C_i(p_i))\\
&   \text{subject to}  & \forall n\in[N], i \in [I],\\
&\lambda:& \sum_{i=1}^I p_i=\sum_{i=1}^I(\sum_{n=1}^N (d_{i,n}-g_{i,n})+e_i),\\
&\mubf:& \Sbf (\sum_{n=1}^N (\gbf_n-\dbf_n)+\pbf-\ebf) \le \fbf,\\
&\underline{\rho}_{in}:&    \max\{\underline{d}_{in}, g_{in}-\overline{c}_{in}\} \le d_{in}, \\ 
&\overline{\rho}_{in}:&  d_{in} \le \min\{\overline{d}_{in},g_{in}+\underline{c}_{in}\}.
\end{array}
\eeq
 KKT conditions of the optimization \eqref{eq: SS_CCDiPro} gives
 \beq
-V_{in}(d^\star_{in})+\lambda^\star-\Sbf_i^\top \mubf^\star+\overline{\rho}_{in}^\star-\underline{\rho}_{in}^\star=0,
\label{eq:KKTSW}
 \eeq
where $\star$ indicates the optimal solution. $\overline{\rho}_{in}^\star\geq 0, \underline{\rho}_{in}^\star \geq 0$, $\Sbf_i\in \mathbb{R}^{L }$ is the $i$-th column of the shift factor matrix $\Sbf$, and $V_{in}(x):=\frac{d}{dx} U_{in}(x)$. Replace in LMP with definition $\pibf: =\mathbf{1}\lambda^\star-\Sbf^\top \mubf^\star$, (\ref{eq:KKTSW}) becomes
 \beq
\pi_i+\overline{\rho}_{in}^\star-\underline{\rho}_{in}^\star=V_{in}(d^\star_{in}).
\label{eq:KKTSW1}
 \eeq
\underline{When $\overline{\rho}_{in}^\star =\underline{\rho}_{in}^\star =0$}, $d^\star_{in}(\pi_i)=V_{in}^{-1}(\pi_i)$ from \eqref{eq:KKTSW1}.

\underline{When $\overline{\rho}_{in}^\star >0$}, we have $d_{in}^\star  = \min\{\overline{d}_{in},g_{in}+\underline{c}_{in}\}$ and $\underline{\rho}_{in}^\star =0$ from the complementarity slackness condition. So \eqref{eq:KKTSW1} becomes
 \beq
\pi_i+\overline{\rho}_{in}^\star=V_{in}(\min\{\overline{d}_{in},g_{in}+\underline{c}_{in}\}).
\label{eq:KKTSW2}
 \eeq
Known that the prosumer utility function is assumed to be concave and continuously differentiable. We have 
 $$V_{in}^{-1}(\pi_i) \geq V_{in}^{-1}(\pi_i+\overline{\rho}_{in}^\star) = \min\{\overline{d}_{in},g_{in}+\underline{c}_{in}\}.$$

\underline{Similarly, when $\underline{\rho}^\star _{in}>0$}, we have $d_{in}^\star =\max\{\underline{d}_{in}, g_{in}-\overline{c}_{in}\}$, $\overline{\rho}_{in}^\star =0$, and $$V_{in}^{-1}(\pi_i) \le V_{in}^{-1}(\pi_i-\underline{\rho}_{in}^\star) = \max\{\underline{d}_{in}, g_{in}-\overline{c}_{in}\}.$$ 

To sum up, we find the optimal consumption of prosumer $n$ at transmission network bus $i$ has 
\begin{align}
\begin{aligned}\label{eq:dstar}
d^\star_{in}(\pi_i)=\min\{{\overline{d}}_{in},g_{in}+\underline{c}_{in}, \max\{&V_{in}^{-1}(\pi_i),\\
&\underline{d}_{in},g_{in}-\overline{c}_{in}\}\},\
\end{aligned}
\end{align}
which equals \eqref{eq:d*}. Here is the reason:
\begin{itemize}
    \item When \underline{$\sum\limits_{n=1}^{{\cal N}_m} g_{in}  \le \sum\limits_{n=1}^{{\cal N}_m} h_{in}(\pi_i) -\underline{C}_m$,} from \eqref {eq:AccessPropsumerLB} we have $\underline{c}_{in}: =
    -g_{in}+h_{in}(\underline{\xi}_m)$. From \eqref{eq:LBxiPi}\eqref{eq:h}, we know $$h_{in}(\underline{\xi}_m)\le h_{in}(\pi_i)\Rightarrow  d^\star_{in}(\pi_i)= g_{in}+\underline{c}_{in}=h_{in}(\underline{\xi}_m)$$ in \eqref{eq:dstar} equals \eqref{eq:d*}.  

    \item When \underline{$\sum\limits_{n=1}^{{\cal N}_m} g_{in} \geq \sum\limits_{n=1}^{{\cal N}_m}h_{in}(\pi_i)+\overline{C}_m$}, from \eqref {eq:AccessPropsumerUB} we have $\overline{c}_{in}: =
    g_{in}-h_{in}(\overline{\xi}_m)$. From \eqref{eq:UBxiPi}\eqref{eq:h}, we know $$h_{in}(\overline{\xi}_m)\geq h_{in}(\pi_i)\Rightarrow  d^\star_{in}(\pi_i)= g_{in}-\overline{c}_{in}=h_{in}(\overline{\xi}_m)$$ in \eqref{eq:dstar} equals \eqref{eq:d*}.

    \item In other cases, we similarly can show $$d^\star_{in}(\pi_i) = h_{in}(\pi_i)=d^*_{in}(\pi_i).$$
\end{itemize}

So, the net production of the prosumer equals \eqref{Q-P}, which aligns the bid/offer curve of the prosumer $n$ at $i$.  Therefore, the social welfare $\mbox{\sf SW}_{\mbox{\rm\tiny DERA}}$ is the optimal value of \eqref{eq: SS_CC}.

By summing up DERA surplus in the objective of  \eqref{eq:DERAsurplus_LnGPCC} and the prosumer surplus from the left-hand side of \eqref{eq:CompetiCons}, we can get the formulation for $ S_{\mbox{\tiny DERA}}$, which is
\begin{align}
\begin{aligned}
S_{\mbox{\tiny DERA}}&=\sum_{i=1}^I \big(\sum_{n=1}^N\omega_n^*-\pi_i (d^{\star}_{in}-g_{in}) \\
&~~~~~~~~+\sum_{n=1}^N (U_{in}(d^{\star}_{in})-\omega_n^*)\big), \\
&=\sum_{i=1}^I \sum_{n=1}^N (U_{in}(d^{\star}_{in})-\pi_i (d^{\star}_{in}-g_{in})).
\end{aligned}
\end{align}\hfill\QED

 \vspace{-0.3cm}
\subsection{Proof of  Theorem~\ref{thm:MarketEfficiency}  }\label{sec:MEproof}

With the prosumer access limits $\overline{c}_{in}$ and $\underline{c}_{in}$ defined in \eqref{eq:AccessPropsumerUB}\eqref{eq:AccessPropsumerLB}, the bidding curve of DERA \eqref{Q-P} is the direct sum of the prosumers' bidding curve (\ref{Q-P_prosumer}). So \eqref{eq: SS_CC} is the market clearing problem when prosumers directly participate in the wholesale market with the bid/offer curve (\ref{Q-P_prosumer}). That way, $\mbox{\sf SW}_{\mbox{\rm\tiny Direct}}$ equals the optimal value of \eqref{eq: SS_CC}, which equals $\mbox{\sf SW}_{\mbox{\rm\tiny DERA}}$. 

By summing up all prosumers' optimal surplus, which is the optimal value of \eqref{eq:ProfitMaxDirect}, over all buses and PoAs, we get $S_{\mbox{\rm\tiny PRO}}$ and it equals $ S_{\mbox{\tiny DERA}}$. \hfill\QED
\subsection{Additional simulation results}
\begin{figure}
\centering
  \scalebox{0.43}{\includegraphics{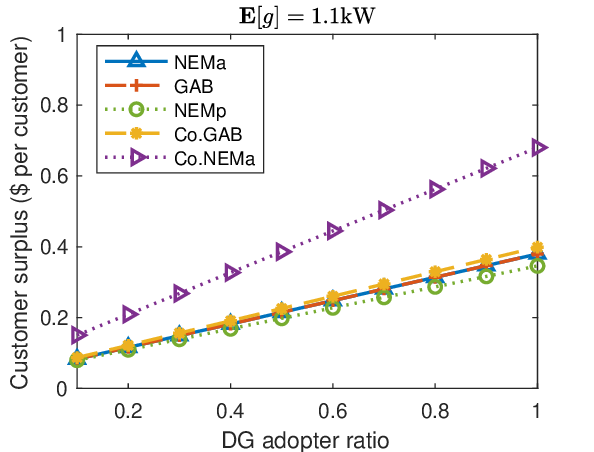}}  \scalebox{0.43}{\includegraphics{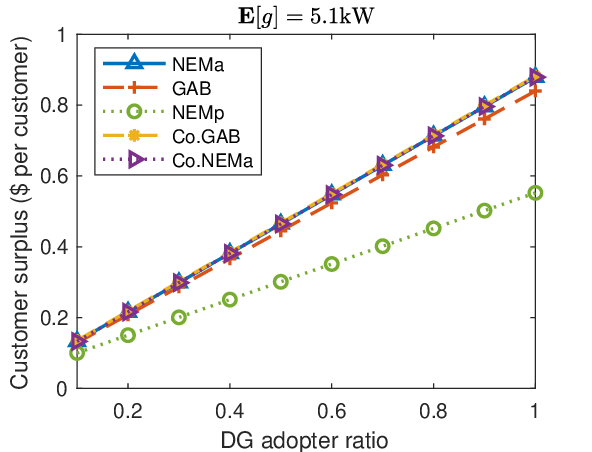}}
 \scalebox{0.44}{\includegraphics{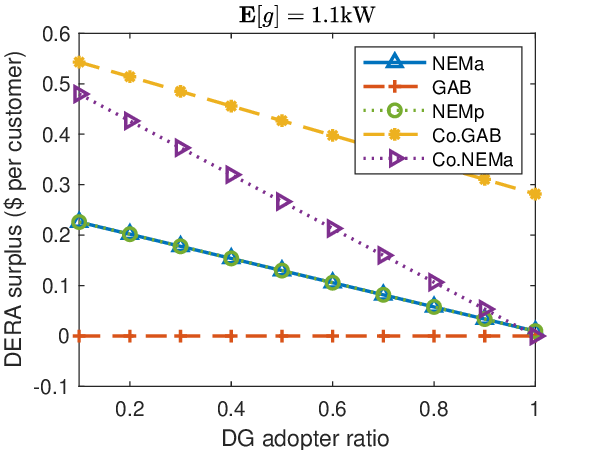}}\scalebox{0.44}{\includegraphics{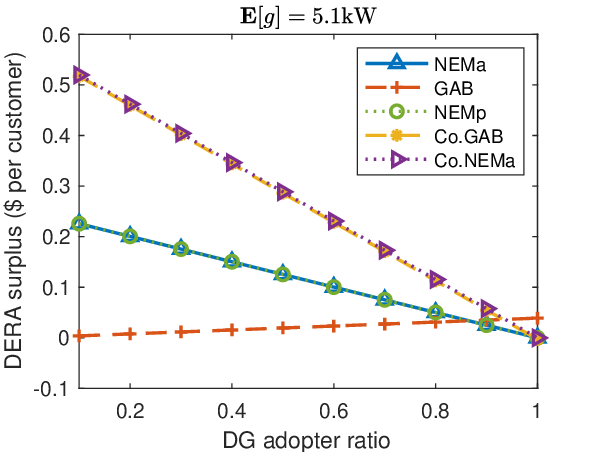}}
 \vspace{-0.5cm}
\caption{Expected surplus distributions vs. network access ratio. (Top: expected customer surplus; bottom: expected DERA surplus.) }
\label{fig:DS2}
\end{figure}

\begin{figure}
\centering
  \scalebox{0.43}{\includegraphics{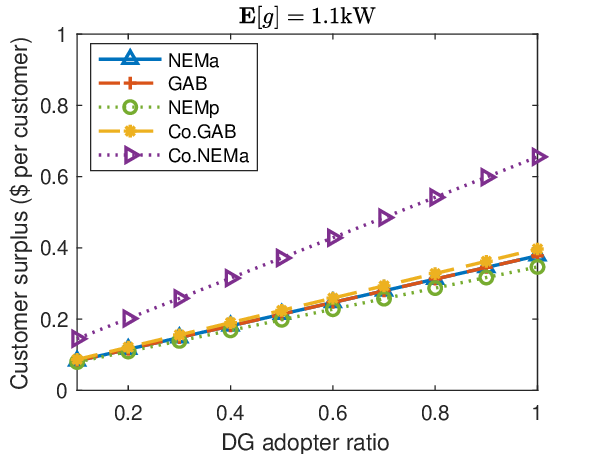}}  \scalebox{0.43}{\includegraphics{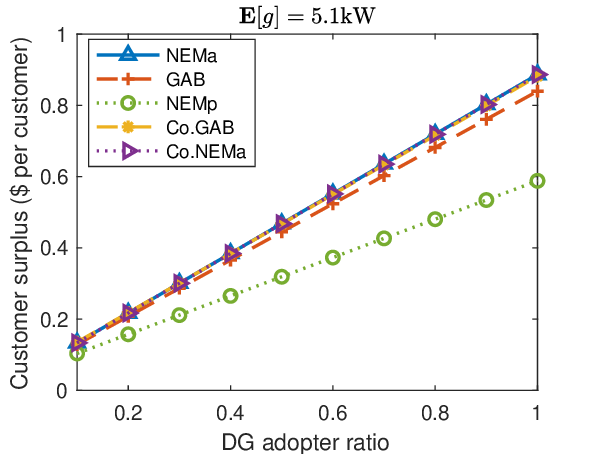}}
 \scalebox{0.44}{\includegraphics{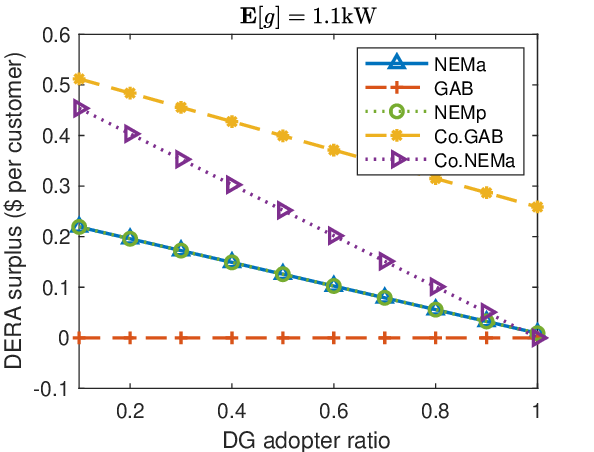}}\scalebox{0.44}{\includegraphics{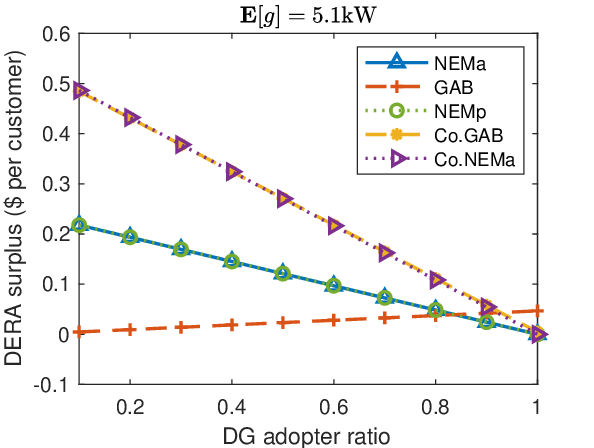}}
 \vspace{-0.5cm}
\caption{Expected surplus distributions vs. network access ratio. (Top: expected customer surplus; bottom: expected DERA surplus.) }
\label{fig:DS3}
\end{figure}

Our method and simulation results are strongly dependent on the mean value of LMP, while the distribution has less influence. In the simulation results of Fig.~\ref{fig:DS}, LMP $\pi$ was modeled as a Gaussian random variable with a mean of \$0.05/kWh and a standard deviation (STD) of \$0.01/kWh. Here, we further show simulation results with a larger STD in Fig.~\ref{fig:DS2}, where LMP is a Gaussian random variable with a mean of \$0.05/kWh and a standard deviation (STD) of \$0.05/kWh. To catch price spike scenarios, we also simulated LMP as a lognormal distribution in Fig.~\ref{fig:DS3}. The lognormal distribution is parameterized with a mean of the underlying normal distribution set to $\mu = \log(0.05)$ and a standard deviation of 0.55. Simulation results in Fig.~\ref{fig:DS2} and Fig.~\ref{fig:DS3} are aligned with Fig.~\ref{fig:DS}.

\subsection{Details about Long-Run Competitive Equilibrium}\label{sec:LREQ}

Here we add details for derivations and parameters in the long-run equilibrium analysis. The long-run equilibrium for single-interval aggregation provides insights into the results in the main text for the multi-interval aggregation.

\subsubsection{Model of long-run competitive equilibrium}

Denote $N$ as the number of aggregated prosumers. In the simulation, we have $N=50$ for each DERA. With the quadratic utility of homogeneous prosumer parameterized by $\alpha$ and $\beta$ in \eqref{eq:UtilityF}, the profit of the $i$-th DERA defined in \eqref{eq:DERAsurplus_LnGPCC} is 
\beq \label{eq:DERAprofit}
\Pi_i(\underline{C}_i) = -\frac{\beta(\underline{C}_i + G_i)^2}{2N} + \alpha (\underline{C}_i+G_i) - \pi\underline{C}_i-{\cal K}_i,
\eeq
where $G_i$ is the aggregated DG generation of DERA $i$ and ${\cal K}_i$ is the competitive benchmark for aggregated prosumers. $\overline{C}_i$ and $\underline{C}_i$ are distribution network injection and withdrawal accesses limits. Here, we only derive the case for network withdrawal access and the case for injection access can be similarly computed.

In the competitive market setting for the distribution network access auction, the network withdrawal access price $\underline{\lambda}$ is assumed to be exogenous. DERA $i$ conducts its profit maximization by $$\underset{\underline{C}_i\geq 0}{\rm maximize} ~ \Pi_i(\underline{C}_i) - \underline{\lambda}\cdot \underline{C}_i.$$ Similarly, DSO's optimization is  $$\underset{\underline{P}\geq 0}{\rm maximize} ~  \underline{\lambda} \cdot\underline{P} - J(\underline{P}),$$ where  $J(\underline{P}): =\frac{1}{2}b \underline{P}^2+a\underline{P}$ is defined as the cost function of DSO for providing the withdrawal access $\underline{P}$ of the distribution network at a certain PoA. For simplicity, we ignore the distribution network reliability constraints in DSO's optimization. 

Denote $S$ as the total number of homogeneous DERAs. Denote the equilibrium price and access allocations as $(\lambda^{\star}, (\underline{C}_i^{\star})_{i \in [S]})$.  We have 
\beq \label{eq:ClearingCondi}
\sum_{i=1}^S \underline{C}_i^{\star} = \underline{P}^{\star},
\eeq
showing the total network withdrawal access is partitioned to individual DERAs. The optimality conditions for these optimizations of DERA and DSO give the first condition for the long-run competitive equilibrium:

(i) The marginal benefit of DERA equals the marginal cost of DSO for providing the distribution network access, \ie 
 \beq \label{eq:condi1}
 \underline{\lambda}^{\star} = \frac{\partial J}{\partial \underline{P}}= b \underline{P}^{\star}+a =  \frac{\partial \Pi_i}{\partial\underline{C}_i} = \alpha - \pi- \frac{\beta}{N} (\underline{C}_i^{\star}+G_i).
 \eeq

The second condition for long-run equilibrium gives:

(ii) all DERAs have profits equal to zero, \ie 
\beq \label{eq:condi2}
\Pi_i(\underline{C}^{\star}_i) - \underline{\lambda}^{\star} \underline{C}^{\star}_i=0.
\eeq

Solve equations \eqref{eq:DERAprofit}\eqref{eq:ClearingCondi}\eqref{eq:condi1}\eqref{eq:condi2}, we find the long-run competitive equilibrium
\beq \label{eq:LREQ}
\underline{C}_i^{\star} =\sqrt{\gamma_i/ \psi}, ~K^{\star} =\frac{2\psi\sqrt{\gamma_i/ \psi} +\beta-b}{2a\sqrt{\gamma_i/ \psi}},
\eeq
where $\gamma_i := \alpha G_i-0.5\beta G_i^2/N-{\cal K}_i$ and $\psi:=-\beta/2N$.

The conditions for the existence of long-run competitive equilibrium are $\gamma_i<0$ and $2\psi\sqrt{\gamma_i/ \psi} +\beta-b \geq 0$.   
    
Interestingly, the wholesale LMP does not influence the long-run equilibrium in \eqref{eq:LREQ} because the linear cost/benefit induced by LMP can be completely cancelled by the marginal pricing at competitive equilibrium. 

\subsubsection{Single-interval long-run competitive equilibrium simulation}
We simulated long-run competitive equilibrium for the single interval aggregation by assuming 200 DERAs existed at the beginning and computed the expected number of surviving DERAs in the long run. For simplicity, we assume homogeneous DERA with the same expectation of BTM DG generation. We sampled 10,000 random scenarios of BTM DG. Same as \cite{ChenBoseMountTong23DERA}, the cost function of DSO when providing distribution network access was assumed to be the sum of quadratics, $\frac{1}{2}bx^2+ax$ with  $a=\$0.009/\mbox{kWh}, b=\$0.0005/(\mbox{kWh})^2$ for both the injection and withdrawal access at all PoAs. 

\begin{figure}[htbp]
    \centering
     \vspace{-0.3cm} 
    \includegraphics[scale=0.37]{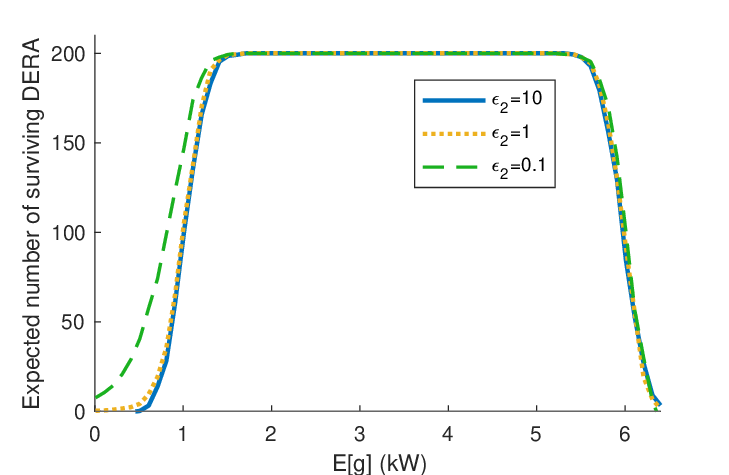}\includegraphics[scale=0.37]{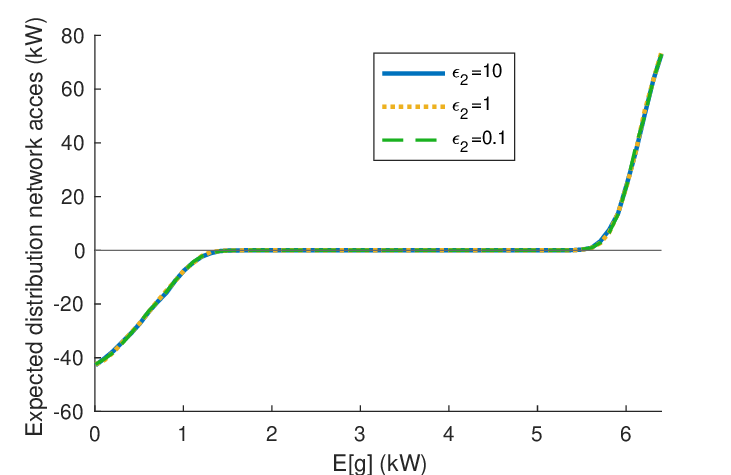}
     \vspace{-0.8cm}
    \caption{Long-run competitive equilibrium. (Left: expected number of surviving DERA; right: expected distribution network net injection access of DERA, whose negativity represents withdrawal access.)}
     \vspace{-0.3cm}
    \label{fig:LREQ_single}
\end{figure}

Three observations were drawn from empirical results in Fig.~\ref{fig:LREQ_single}. First, when the expected BTM DG was about 2-5 kW, all initial 200 DERAs survived and the expected net injection access equals zero. It's because DERA internally balanced customer demands with BTM DG, thus relying less on competing for the injection or withdrawal accesses. Second, with smaller expected BTM DG, homogeneous DERAs competed for the withdrawal access to the distribution network, and less than 10 DERAs survived in the long run; with larger expected  BTM DG, DERAs competed for the injection access and less than 3 DERAs survived. Fewer DERAs survived when competing over the injection access because NEM X credited DG imports well, making DERA survival more challenging under high DG generations. Third, with smaller $\epsilon_2$ for the DSO's cost scaling factor, the DERA payment to the network access was lower, thus more DERA survived in the green dash curve.  

 \subsubsection{Multi-interval long-run competitive equilibrium simulation} 
\begin{figure}[htbp]
    \centering
     \vspace{-0.3cm}
    \includegraphics[scale=0.37]{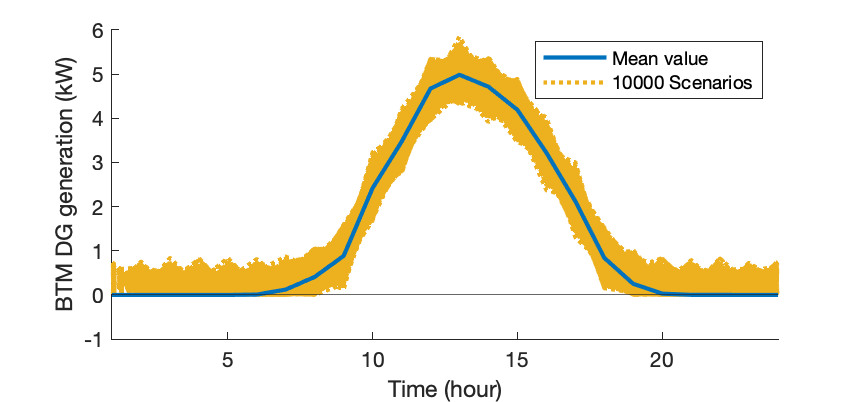}
    \caption{Mean and 10,000 scenarios of BTM DG generation from prosumer.}
     \vspace{-0.3cm}
    \label{fig:Solar}
\end{figure}

By adding the 24-hour time dimension to the network access and BTM DG generation, we can extend the derivation of \eqref{eq:DERAprofit}\eqref{eq:ClearingCondi}\eqref{eq:condi1}\eqref{eq:condi2}  from single-interval long-run equilibrium to multi-interval long-run equilibrium. Note that the number of DERA $K$ is still a scalar applied to all 24 hours. We include the simulation setting and results for the multi-interval aggregation in Sec.~\ref{sec:EQ_MultiT}. The solar scenarios used in the simulation are presented in Fig.~\ref{fig:Solar}.

\end{document}